\documentclass[letterpaper,twocolumn,10pt]{article}

\usepackage{usenix,epsfig}
\usepackage{floatrow}
\usepackage{verbatim}
\usepackage{balance}
\usepackage[utf8]{inputenc}
\usepackage[cmex10]{amsmath}
\usepackage{amssymb}
\usepackage{color}
\usepackage{amsfonts}
\usepackage{amsthm}
\usepackage{booktabs}
\usepackage{algorithmic}
\usepackage{algorithm}
\usepackage{array}
\usepackage{url}
\usepackage[font=small]{caption}
\usepackage[font=small]{subcaption}
\usepackage{cite}
\usepackage{subfig}
\newtheorem{example}{Example}

\usepackage[compact]{titlesec}
\usepackage{color}
\usepackage{multirow}
\usepackage{balance}

\newcommand\co[1]{}

\newcommand\FF{\mathbb{F}}
\newcommand\E{\text{E}}

\newcommand\Var{\text{Var}}
\newcommand\bv[1]{\mathbf{#1}}

\mathchardef\d="2D

\usepackage{graphicx}
\usepackage{ifpdf}
\ifpdf
\fi

\begin{document}

\title{\Large \bf The CORE Storage Primitive: Cross-Object Redundancy for\\ Efficient Data Repair \& Access in Erasure Coded Storage \thanks{This work is supported by Singapore's A*Star Grant 102 158 0038 \& National Research Foundation Grant NRF-CRP2-2007-03.}}

\author{
  {\rm Kyumars Sheykh Esmaili, Lluis Pamies-Juarez, Anwitaman Datta}\\
    Nanyang Technological University, Singapore\\
  \{kyumarss,lpjuarez,anwitaman\}@ntu.edu.sg
} 
\date{} 
\maketitle

\thispagestyle{empty}

\begin{abstract}
Erasure codes are an integral part of many distributed storage systems aimed
at Big Data, since they provide high fault-tolerance for low overheads.
However, traditional erasure codes are inefficient on reading stored data
in degraded environments (when nodes might be unavailable), and on
replenishing lost data (vital for long term resilience). Consequently,
novel codes optimized to cope with distributed storage system nuances are vigorously being researched. In this paper, we
take an engineering alternative, exploring the use of simple and mature
techniques -- juxtaposing a standard erasure code with RAID-4 like parity. We carry out an analytical study to determine the efficacy of this approach over traditional as well as some
novel codes. We build upon this study to design CORE, a general storage primitive that we integrate into HDFS. We benchmark this
implementation in a proprietary cluster and in EC2. 
Our experiments show that compared to traditional erasure codes, CORE uses
50\% less bandwidth and is up to 75\% faster while recovering a single failed node, while the gains are respectively 15\% and 60\% for double
node failures.
\end{abstract}


\section{Introduction}\label{s:intro}
In order to  meet the conflicting needs of high fault-tolerance and low storage overhead, erasure codes are increasingly being embraced for distributed storage systems\footnote{Scaling-out (horizontal scaling) has become a norm to deal with very large scale storage, and this work is thus focused on distributed storage system architectures. We do not elaborate further the usual motivations and compulsions for such a system design choice.} aimed to store high volumes of data. Traditional erasure codes have mostly been designed to optimize the performance of communication-centric applications, and are not necessarily amenable to the needs of storage systems. Some such desirable properties include efficient replenishment of lost redundancy (repair) following the failure of some system components; and efficient access of data while the system is yet to complete remedial actions following such failures (degraded reads/access). To that end, there has been tremendous interest in both coding theory and storage systems research communities to build new erasure codes with good repairability properties, as well as building robust storage systems leveraging on the novel codes (for instance, Windows Azure Storage using Local Reconstruction Codes). In this paper we explore an alternate design, looking at an instance of product codes \cite{productcode}. A traditional erasure code is first applied on individual data objects, followed by the creation of RAID-4 like parity over erasure encoded pieces of different objects, creating \underline{c}ross-\underline{o}bject \underline{re}dundancy.  This results in high fault tolerance (provided by the traditional code) and cheap repairs (provided by the parity code). The approach is simple, and based on mature techniques that have long been used as stand-alone approaches (these are desirable for practical and implementation considerations), yet it achieves very good (less communication \& computation) repairability and degraded data access under many fault-conditions. We accordingly build the \emph{CORE storage primitive} as a general purpose, block level, fault-tolerant, data storage layer that can be readily integrated into distributed file systems relying on an underlying block level storage, providing significant performance boost. We integrate CORE into Hadoop Distributed File System (HDFS), and benchmark it over a wide range of system configurations, comparing it with state-of-the-art alternatives \cite{hdfsraid,azureec} to demonstrate its efficacy.

CORE builds upon our recent work \cite{ApsysRGC} where we made a simple
observation - by introducing a RAID-4 like parity over a small set of erasure
encoded pieces, it is possible to achieve significant reduction in the
expected cost to repair lost redundancy. Moreover, since these extra parities are relatively-small, the fault tolerance of the resulting system is only marginally lower than what is achieved with optimal
(maximum distance separable, or MDS) codes --i.e. Reed-Solomon codes.\footnote{This suboptimal fault-tolerance is expected from any code aiming to
reduce the costs of repairs. The trade-offs between repair costs and fault-tolerance have recently been determined by Gopalan et al. in~\cite{locbound}.}

\begin{figure}%
\includegraphics[trim=3cm 7cm 3cm 6cm,width=\columnwidth]{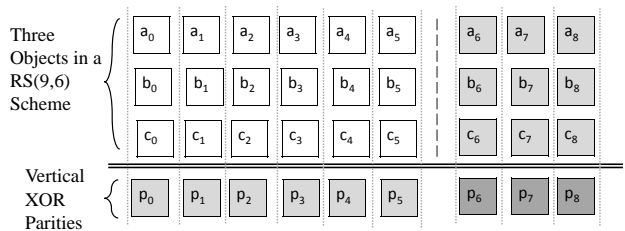}%
\caption{An example illustrating the basic idea of CORE}%
\label{fig:example1}%
\end{figure}

In Figure~\ref{fig:example1} we show an example to elaborate the basic idea on which CORE is built. Consider three objects ($a$,
$b$, $c$), each comprising of 6 blocks. Each of these objects are first individually encoded using a (9,6) Reed-Solomon code. Note that each row represents an object along with its
three parity blocks (depicted in gray). Additionally, a simple parity
check (i.e. an XOR) is computed over each column's blocks and thereby a new
row is added at the bottom of the matrix. In this example, this extra
row increases the storage overhead by 33\%. However, as we will show later
(e.g., Figure~\ref{f:resil14}), CORE's parameters can be adjusted to operate at reasonable overheads, even while achieving very good fault-tolerance as well as repairability. In particular, for equivalent storage overhead, CORE's performance benefit is significantly better than the state-of-the-art Locally Reconstruction Codes used in Azure \cite{azureec}, while, CORE achieves fault-tolerance (ignoring repairs) comparable to optimal MDS erasure codes for an acceptable 20\% more storage overhead.
	
The main advantage that the vertical parities introduce in CORE is the increased efficiency of repairs. Fewer blocks are needed to carry out a repair. Furthermore, repair related computation is cheap (a simple XOR operation, compared to the expensive RS decoding procedure). In the example of Figure~\ref{fig:example1}, repairing any single failure would require XORing 3 blocks.\footnote{This process can furthermore be pipelined across the vertical nodes to avoid the delay required in downloading the three blocks at the node (say, due to bandwidth bottlenecks) where repair is carried out. However the current CORE implementation does not do so, since it would require changes to the underlying HDFS layer, which we deliberately avoided for achieving CORE's ready portability with HDFS.}
Apart from the repairability benefits, CORE's vertical parities also improve fault tolerance. For instance, in Figure~\ref{fig:example1}, an object (i.e. a row) with more than three failures can be still recovered with the help of vertical parities. This improvement is however, not optimal in terms of the additional storage overhead, as the primary purpose of the vertical parities is to enhance the repairability aspect.
The low repair cost in CORE also naturally translates into better degraded reads.

\color{black}{}

The main contributions of this work are as
follows:

\begin{itemize}
\item While this work builds on the preliminary observation
\cite{ApsysRGC} (and more generally, on the idea of product codes
\cite{productcode}), from the theoretical perspective, the main
contribution of this paper is a rigorous analysis and confirmation
of the intuitions advanced in \cite{ApsysRGC}, and comparison with
traditional erasure codes as well as novel storage-centric code,
namely local reconstruction code \cite{azureec} used in Azure.
\item  A more practical contribution is from the systems research perspective,
where we implement the ideas to build the general purpose block
level storage primitive CORE, and integrate it with a popular
distributed file system, HDFS (our implementation is available at \cite{TheCOREurl}).
\item In the process, we identify a few ways to optimize the
HDFS-RAID \cite{hdfsraid} implementation, on which we build CORE.
\item We design novel algorithms to understand the failure patterns
and adaptively exploit the better flexibility afforded by CORE's code design,
in order to achieve fast and cheap repairs. The implementation is
meticulously tested and benchmarked in a proprietary cluster as well
as on EC2.
\end{itemize}
To repair single failures, CORE consumes 50\% less bandwidth and is between 43\% to 76\% faster  compared the classic erasure code.
In case of double failures and in the worst case scenario --when both failed blocks belong to the same file-- it consumes 16\% less bandwidth and is  13\% to 59\% faster. We thus hope that CORE's design and analysis is not only academically interesting, but the performance boost it achieves, and the fact that it is based on a
simple composition of existing mature techniques (instead of relying on proprietary
techniques or other untested novel approaches) makes it a serious candidate
for wide-scale adoption.

\section{Related Work}
\label{s:relwork}

Erasure codes have long been explored as a storage efficient alternative to replication for achieving fault-tolerance \cite{codingvsrepl} in the peer-to-peer (P2P) systems literature, and have led to numerous prototypes, e.g., OceanStore \cite{OceanStore} and TotalRecall \cite{totalrecall} to name a few. In recent years erasure codes have gained traction \cite{liber8tion} even in main-stream storage technologies such as RAID \cite{RAID}. The ideas from RAID systems are in turn permeating to Cloud settings \cite{khan2012rethinking,diskreduce2}, and erasure codes have become an integral part of many proprietary file systems used in data-centers \cite{azurestorage,warehouseface}, as well as open-source variants \cite{hdfsraid}.

With the proliferation of erasure codes in storage-centric applications,
there has been a corresponding rise in the exploration of novel erasure codes
which cater to the nuances of distributed storage systems. Specific aspects
that have been investigated in designing such new coding techniques include: (i) \emph{efficient degraded data access} \cite{pyramid,azureec}, (ii)
\emph{good repairability}\cite{surveyreplong,regeneratingsurvey} by either
combining standard codes \cite{hierarchicalcodes,GridCodes,ApsysRGC},
applying network coding techniques \cite{Kermarrec,Schum,regeneratingsurvey},
or designing completely new codes with lower repair fan-in
\cite{OD,ODITW,papailiopoulos2012locally,Xorbas}, and (iii) \emph{fast creation} of
erasure coded redundancy \cite{icdcit13,rapidraid}.

Despite the plethora of works investigating novel erasure codes,
most existing distributed file systems using erasure codes do so by
adapting traditional erasure codes. Microsoft's Windows Azure Storage
\cite{azurestorage} is a prominent exception which uses an optimized
version of Pyramid codes \cite{pyramid} called Local Reconstruction
Code (LRC) \cite{azureec}. Some recent academic prototypes - NCFS
\cite{NCFS} and \cite{COREHK}\footnote{Coincidentally, \cite{COREHK} uses the same name, CORE, for \underline{co}llaborative \underline{re}generation.} likewise explore the
feasibility of applying network coding techniques for repairing lost
data. The latter systems do not address the issue of degraded reads. In contrast to these systems based on proprietary and novel erasure coding techniques with significant system design complexity, CORE composes two mature techniques (standard erasure codes and RAID-4 like parity) while achieving very good repairability and degraded read performance. This makes CORE suitable for ready integration with many block based storage/file systems, and its simple design makes it amenable to third party reimplementations.

\section{Background} 
\label{s:background}
We next provide some background on what are erasure codes and how they are used in distributed storage systems, followed by a discussion on local repairability which has come to the fore in the design of novel storage-centric erasure codes. Finally, we discuss the code used in Azure system, which, to the best of our knowledge, was the first deployment of repairable codes in a large-scale commercial cloud storage system.  
\subsection{Classic Erasure Codes}
\label{s:bc}

Traditionally, large data objects have been stored
by splitting them into blocks of (say) size $q$ bits, which are then replicated
across multiple storage nodes. In contrast, an ($n,k$) erasure code takes $k$
different data blocks of size $q$, and computes $m=n-k$ parity blocks of the
same size, each to be stored in a different storage node. Then, in the event
of disk failures, the $k$ original blocks can be reconstructed by collecting
and decoding a subset of $k' \geq k$ blocks out of the total $n$ stored blocks.

Consider that the vector $\bv o=(o_1,\dots,o_k)$ denotes a
data object composed of $k$ blocks of $q$ bits each. That is, each block
$o_i$ is a string of $q$ bits. The encoding operations are performed using
finite field arithmetic where the two bits $\{0,1\}$ form a
finite field $\mathbb{F}_2$ of two elements, while $o_i$ likewise belongs to the
binary extension field $\mathbb F_{2^q}$ containing $2^q$ elements. Then, the
encoding of the object $\bv o$ is a linear transformation defined by a
$k\times n$ generator matrix $G$ such that we can obtain an $n$-dimensional
codeword ${\bv c}=(c_1,\dots,c_n)$ of size $n\times q$ bits by applying the
linear transformation $\bv c = \bv o \cdot G$. A code with such a generator
matrix $G$ is usually referred to as an ($n,k$)-code. When the generator
matrix $G$ has the form $G=[I_k,G']$ where $I_k$ is the identity matrix and
$G'$ is a $k\times m$ matrix ($m=n-k$), the codeword ${\bv c}$ becomes ${\bv
c}=[{\bv o},{\bv p}]$ where $\bv o$ is the original object, and ${\bv p}$ is
a parity vector containing $m\times q$ parity bits. The code is then said to
be \emph{systematic}, in which case the $k$ parts of the original object remain
unaltered after the coding process.
We want to
note that the main advantage of systematic codes is that the original data
$\bv o$ can be accessed without requiring a decoding process, by just reading
the systematic blocks of $\bv c$.

The above encoding process stretches the original data by a factor of $n/k$
(ratio known as the \emph{stretch factor}), occupying $n/k$
times more storage space than the size of the original object. By choosing
a suitable code with a stretch factor satisfying $n/k<r$, significant storage
space savings can be achieved in comparison to a system using $r$ replicas.
Finally, an \emph{optimal erasure code} in terms of the trade-off between
storage overhead and fault tolerance is called a \emph{maximum distance
separable} (MDS) code, and has the property that the original object $\bv o$
can be reconstructed from any $k$ out of the total $n=k+m$ stored blocks (i.e., $k'=k$),
tolerating the loss of any arbitrary $m=n-k$ blocks. The fault-tolerance or
MDS erasure codes has been variously analyzed and compared with replication
\cite{codingvsrepl,totalrecall}, providing guidelines to choose suitable code
parameters $n$ and $k$ for a desired level of resilience under an expected
level of failures of individual storage nodes.

\subsection{Locally Repairable Codes}
\label{s:lrc}
A critical drawback of MDS codes is their high reconstruction cost.
Repairing/reading a single failed block requires to
download an amount of information equivalent to the size of the whole data
object $\bv o$, which is $k$ times larger than the amount of data being repaired/read.

Since repairs and degraded reads are frequent in storage systems, several
recent works \cite{OD,ODITW,pyramid,azureec,papailiopoulos2012locally} have
looked at reducing the number of blocks needed to carry out the
repair/reconstruction of an inaccessible block (which is needed for both access
and repair). Such a property is achieved by introducing `local dependencies'
among encoded blocks, and can be called repair locality. 

Local repairability is achieved when a
block $c_i$ can be expressed as a linear combination of $d$ ($d<k$) other blocks,
$c_i=\alpha_1c'_1+\alpha_2c'_2+\dots+\alpha_dc'_d$, and where $c'_j\in\bv c$
s.t.  $c'_j\ne c_i$, where the coefficients $\alpha_j\in\FF_{2^q}$ have
predetermined values. This local repairability property allows to reduce the
number of blocks accessed and transferred during degraded reads or repairs
from $k$ to $d$, where $d$ can be as small as $d=2$~\cite{OD,ODITW}.
Unfortunately, achieving such code locality leads to poorer fault-tolerance for a given storage overhead in comparison to MDS codes. Hence, the design of such codes poses a
trade-off between three important desirable system properties: (i) high
fault-tolerance, (ii) low storage overhead, and (iii) efficient repairs and
degraded reads.

\subsection{Local Reconstruction Code in Azure}
\label{s:lrcinazure}
Local Reconstruction Code (LRC) \cite{azureec} used in the Azure
system is an instance of Pyramid codes~\cite{pyramid}
optimized to achieve a good trade-off among these desirable
properties. In its simplest form, an ($n,k$) LRC code (for
even $k$s, and $n\geq k+2$) is a code composed of two classic
optimal (MDS) erasure codes: (i) a systematic $(n-2,k)$-code with a
\emph{global} generator matrix $G_g$ of the form $G_g=[I_k,H]$, and (ii) a
systematic $(k'+1,k')$-code with \emph{local} generator matrix $G_l$ of the
form $G_l=[I_{k'},\mathbf{1}_{k'}^\top]$, for $k'=k/2$. Then, the LRC
encoding process consists of splitting the original data vector $\bv o$ into
two equal-sized vectors $\bv o = (\bv o_1, \bv o_2)$ (recall that $k$ is even) and perform three independent encoding operations:
\begin{align*}
\bv c_g &= \bv o\cdot G_g=[\bv o,\bv p_g], \\
\bv c_1 &= \bv o_1\cdot G_l = [\bv o_1,\bv p_1], \\
\bv c_2 &= \bv o_2\cdot G_l = [\bv o_2,\bv p_2].
\end{align*}

\noindent Then, the final codeword of the LRC code $\bv c$ can be obtained by
concatenating the parity vectors of the three previous independent encodings:
$\bv c=[\bv o_1, \bv o_2, \bv p_1,\bv p_2,\bv p_g]$.

The local reconstruction property of this composed code can be shown as
follows. When a single codeword block $c_i\in\bv c$ is missing, it can be
repaired by reconstructing either $\bv o_1$ (if $c_i\in\bv c_1$) or $\bv o_2$
(if $c_i\in\bv c_2$), and regenerating the missing block $c_i$ from it. This
repair mechanism entails transferring $k'=k/2$ redundancy blocks over the
network, a half of the traffic required by a MDS $(n,k)$ erasure code.
However, when the missing block $c_i$ does not belong to $\bv c_1$ or $\bv
c_2$ (or $c_i\in\bv p_g$), the repair cannot exploit
the local reconstruction property and has to repair $c_i$ using the global
code, transferring then $k$ redundancy blocks. It means that LRC codes only
allow to locally repair $2(k'+1)=k+2$ blocks out of the total $n$ blocks in the
codeword. The remaining $n-k-2$ blocks have to be repaired by downloading
$k$ blocks.

\co{in terms of a generator matrix\linebreak
$G=[I_k,(I_2\otimes\mathbf{1}_{k/2})^\top,H]$, where $I_k$ is the
identity matrix, $\mathbf{1}_{k/2}$ is an all-one vector of size $k/2$, and
$H$ is a $k\times(n-k-2)$ matrix, defined as
\[
H = \begin{pmatrix}
\alpha_1^1 & \dots & \alpha_1^{n-k-2} \\
\vdots & \ddots & \vdots \\
\alpha_k^1 & \dots & \alpha_k^{n-k-2} \\
\end{pmatrix},
\]
where $\alpha_1,\dots\alpha_k\in\FF_{2^q}$. Additionally, the $\otimes$
operation in the definition of $G$ denotes the Kronecker product, which for a
$m\times n$ matrix $X$ and an arbitrary matrix $Y$ is defined as
\begin{equation}
X\otimes Y =
\begin{pmatrix}
  x_{11}Y & \cdots & x_{1n}Y \\
  \vdots  & \ddots & \vdots  \\
  x_{m1}Y & \cdots & x_{mn}Y
 \end{pmatrix}.
\label{e:kprod}
\end{equation}
}

\begin{figure}
\centering
\includegraphics[scale=.23]{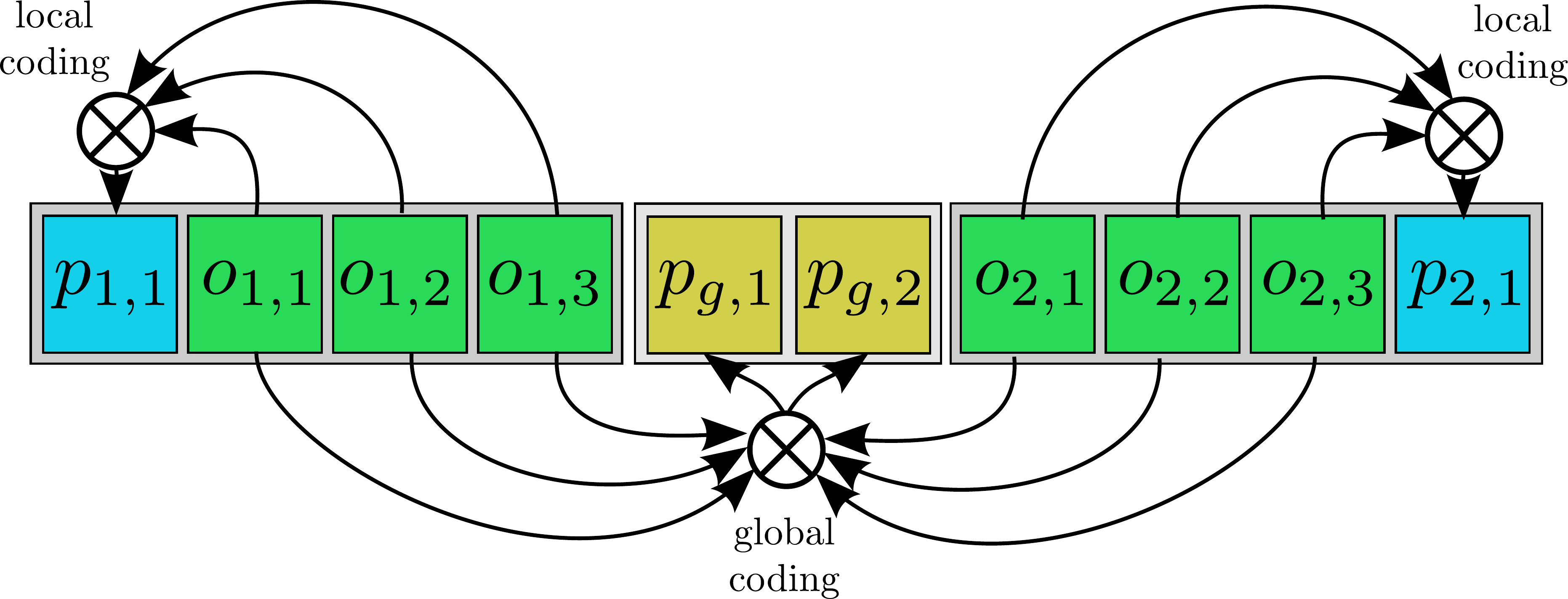}

\caption{Example \hfill of \hfill a \hfill(10,6) LRC \hfill code\hfill
encoding \hfill an object\\ $\bv o=(\bv o_1,\bv o_2)=(o_1,\dots,o_6)$. A
MDS (4,3) erasure code (the \emph{local} code) generates two parity blocks
($p_{1,1}$ and $p_{2,1}$), and another MDS (8,6)-code (the \emph{global}
code) generates the other two parity blocks ($p_{g,1}$ and $p_{g,2}$).}
\label{fig:expyram}
\end{figure}

In Fig.~\ref{fig:expyram} we depict an example of how this (10,6) LRC code
encodes the different systematic blocks from two different sub-objects $o_1$
and $o_2$. Repairing any block from $\bv o$, $\bv p_1$ or $\bv p_2$ can be
done by accessing three other blocks, e.g., $o_{1,2} =
o_{1,1}+o_{1,3}+p_{1,1}$. However, repairing a block from $\bv p_g$ requires
to access six other blocks.

Besides being able to repair single block
failures, LRC can also repair the simultaneous failure of multiple blocks.
LRC can repair any combination of $n-k-2$ failures using the global code; and
can repair up to $n-k$ simultaneous failures by combining local and global
repairs. On an average, repairing any one missing block in LRC requires
$\left(\frac{k+2}{n}\right)\frac{k}{2}+\left(\frac{n-k-2}{n}\right)k = (2kn-k^2-2k)/2n$ blocks.
In Section~\ref{s:core2} we will analyze in more detail the
resiliency of LRC to multiple block failures.

\section{Cross-Object Redundancy}
\label{s:core1}

The reconstruction
locality of LRC used in Azure offers good read performance under degraded system conditions.
However, Pyramid codes~\cite{pyramid} (the framework behind LRC) were
not originally conceived for efficient repairs per say, and as shown in the above example, not all blocks can be efficiently repaired. Although LRC codes can significantly reduce the
repair traffic as compared to MDS codes, there are still
cases where data locality cannot be fully exploited to repair arbitrary
missing blocks.

We next explore how \emph{product codes} \cite{productcode} can achieve good repairability
without compromising either the degraded read performance or the
fault-tolerance of the code. Specifically, by combining a long and a short linear erasure code, we realize
a product code with high fault tolerance (provided by the long
code) and high repair locality (provided by the short code). This is achieved by encoding multiple
already-encoded objects together (or \emph{cross-object} encoding), thus reusing existing encoding/decoding/repair mechanisms already
deployed in a distributed storage system, facilitating an organic
integration of the approach.

\vspace{-2mm}
\begin{example}
Suppose that we have two different data objects $\bv
o_1=(o_{11},o_{12},o_{13})$ and $\bv o_2=(o_{21},o_{22},o_{23})$ to be
encoded with a (5,3) systematic MDS erasure code (with a generator matrix
$G_{\bv o}$). Then, we obtain the codewords:
\vspace{-2mm}
\begin{align*}
\bv c_1 &= \bv o_1\cdot G_{\bv o} = (o_{11},o_{12},o_{13},p_{11},p_{12}), \\
\bv c_2 &= \bv o_2\cdot G_{\bv o} = (o_{21},o_{22},o_{23},p_{21},p_{22}).
\end{align*}

\begin{figure}
\centering
\includegraphics[scale=.23]{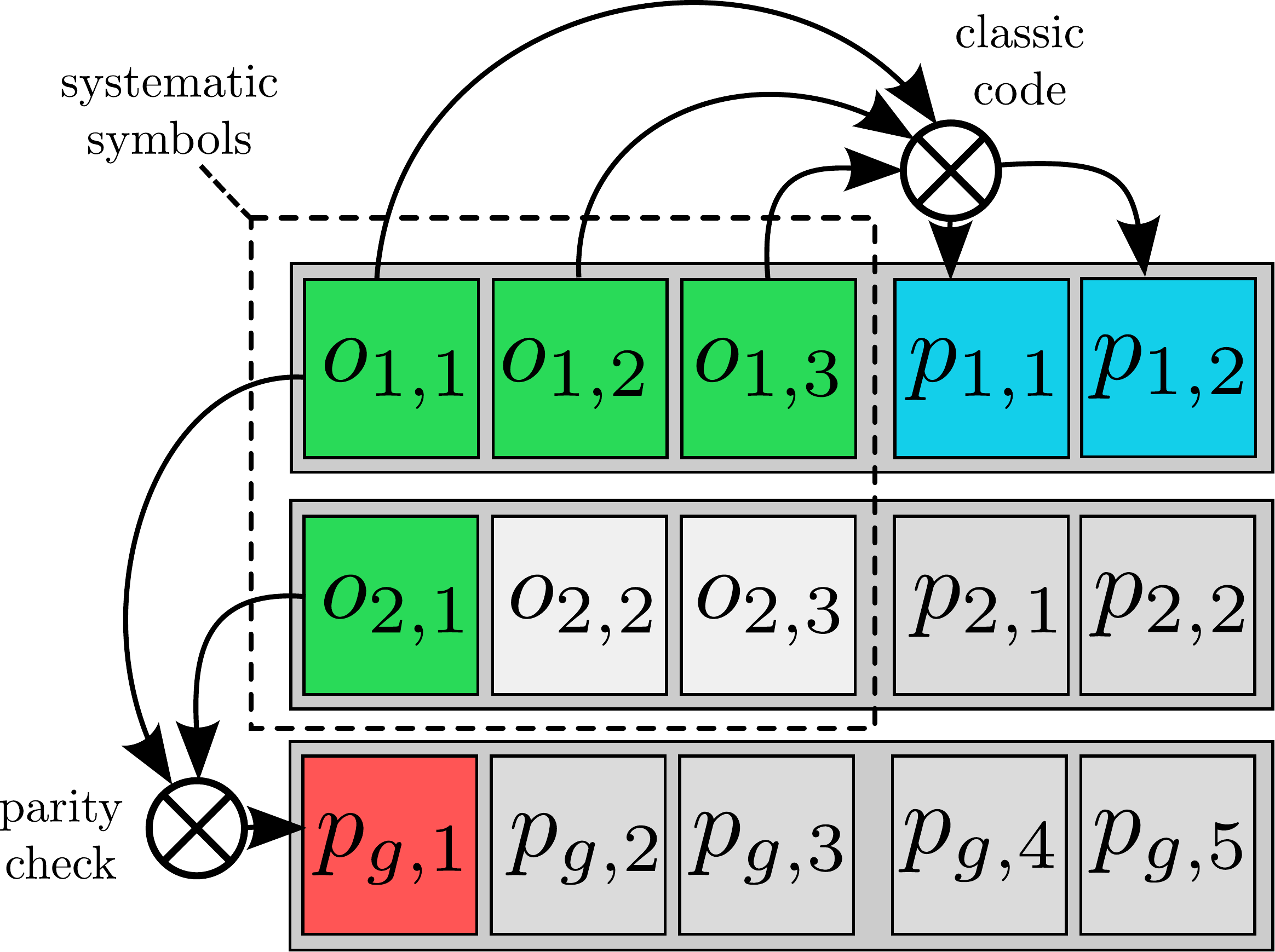}

\caption{Example of a simple product code. The blue parity blocks
are generated using a horizontal (5,3) MDS code whereas the green
blocks are simple parity checks of each column (or a (3,2) code).}
\label{fig:exprod}

\end{figure}

By grouping symbols from $\bf c_1$ and $\bf c_2$ in a per-column basis, we
obtain the set of vectors
$$
\mathcal P=\{(o_{11},o_{21}),(o_{12},o_{22}),(o_{13},o_{23}),(p_{11},p_{21}),(p_{12},p_{22})\}.
$$

\noindent We encode then each vector $x_i\in\mathcal P$ (cross-object encoding) with a
(3,2) systematic code (a simple parity check code, or SPC), with generator
matrix $G_{\bv g}=[I_2,\mathbf{1}_2]$), where $I_2$ is the identity
matrix, and $\mathbf{1}_2$ is a vector with two ones. For each
$x_i\in\mathcal P$ we obtain $\bv p_{g,i}=x_i\cdot G_{\bv g}=[x_i,\bv
p_{g,i}]$, where $\bv p_{g,i}=(\sum x_i)$. The vector with all the
cross-object parity blocks, $\bv p_{g}=(\bv p_{g,1},\dots,\bv p_{g,5})$, contains:
\vspace{-2mm}
$$
\bv p_{g} = (o_{11}\!+\!o_{21},~o_{12}\!+\!o_{22},
o_{13}\!+\!o_{23},~p_{11}\!+\!p_{21},~p_{12}\!+\!p_{22}).
\vspace{-2mm}
$$
In Fig.~\ref{fig:exprod} we depict this two-phase encoding process. Note that
$\bv p_{g}$ can be viewed as the Reed Solomon encoding of the respective
parities of the systematic symbols. We refer to a code with a generator
matrix $G$ that takes a composed data object $\bv o=(\bv o_1, \bv o_2)$ and
encodes it to a codeword $\bv c=\bv o\cdot G=[\bv c_1, \bv c_2, \bv p_g]$, as
the product code of $G_{\bv g}$ and $G_{\bv o}$. It is easy to see how this
example product code repairs any single missing block by using the outer
erasure code $G_\bv g$, e.g., we can repair $o_{1,1}$ using $o_{1,1} =
o_{2,1}+p_{g,1}$. In addition, in case of more than one failure per
``column'', the code still has the opportunity to repair up to two failures
per codeword $\bv c_i$, and up to two failures within the additional parity
vector $\bv p_{g}$.
\end{example}


\noindent\textbf{Definition of CORE's Product Code:} Let $G_{\bv c}$ and $G_{\bv o}$ respectively be the generator matrices of an
$(n_c,k_c)$ and an $(n_o,k_o)$ code. Then, the product
code of $G_{\bv c}$ and $G_{\bv o}$ is a ($n_cn_o,k_ck_o$) linear code with
generator matrix $G=G_{\bv c}\otimes G_{\bv o}$, where the operator $\otimes$
represents the Kronecker product.

In the case of the product code used in CORE, we will consider that
the single parity check (SPC) with generator matrix $G_{\bv c} =
[I_t, {\bf 1}_t]$, i.e., a $(t+1,t)$ MDS erasure code over
$\FF_{2^q}$ is the {\bf vertical} code. For an input $\bv
o=(o_1,\dots,o_t)$, $o_i\in\FF_q$, this code generates a systematic
codeword $\bv c=(c_1,\dots,c_{t+1})=(o_1,\dots,o_t,c_{t+1})$, where
$c_{t+1}=\sum_{i=1}^t o_i$. Since $\FF_{2^q}$ is a binary extension
field the last symbol in the codeword corresponds to the
exclusive-or (XOR) of the $t$ original symbols. It can repair any
single erasure in the codeword by xoring the remaining $t$ symbols.
The inner ({\bf horizontal}) code $G_{\bv o}$ used in CORE is a MDS
($n,k$) erasure code. For the sake of simplicity, we will consider
that it is a ($n,k$) Reed-Solomon code with generator matrix $G_{\bv
o} = [I_k, H]$, where $H$ is a $k\times m$ Vandermonde matrix
(recall that $m=n-k$),
$$
H= \begin{pmatrix}
\alpha_1^0&\dots&\alpha_{1}^{m-1}\\
\vdots&\ddots&\vdots\\
\alpha_k^0&\dots &\alpha_{k}^{m-1}
\end{pmatrix},
$$
\noindent for any $\alpha_i\in\FF_{2^q}$. Then, the CORE's product code is a
linear code that cross-encodes $t$ different data objects using a generator
matrix $G=G_{\bv c}\otimes G_{\bv o}$. We will refer to such a code as a
$(n,k,t)$ CORE product code.

\section{Analysis of CORE's Product Code}
\label{s:core2}

In this section we evaluate CORE's product code in terms of its (i)
fault-tolerance, (ii) repair traffic, and (iii)
efficiency of reading data in degraded situations and compare it with MDS erasure codes and LRC codes.

\subsection{Fault Tolerance}
\label{s:ft}

One way to compare the data reliability of different erasure codes is to
measure the amount of data lost (non repairable objects) when a fraction $p$
of all storage nodes fail simultaneously. In a large distributed storage
system (with thousands of nodes) the average amount of data lost is
equivalent to $N\times\pi$, where $N$ is the total number of stored objects
and $\pi$ is the probability of being able to repair any single object when
each stored block is independently accessible with probability $p$. This probability $\pi$ is used as a metric to quantify the fault-tolerance of a code and is called its \emph{static resilience}.

\vspace{2mm}
\noindent {\bf MDS Erasure Codes:}

\noindent The static resilience of an MDS ($n,k$) erasure code, denoted as $\pi_{E}$, can be measured as the probability that at most
$m=n-k$ redundant blocks are inaccessible, i.e., $\pi_{E} = \Pr(B(n,p)\leq
m)$, where $B(n,p)$ represents a Binomial variate describing the number of
inaccessible blocks in a set of $n$ blocks when nodes are inaccessible with
probability $p$, hence,
\[
\Pr(B(n,p) \leq m) = \sum_{i=0}^m \binom{n}{i} p^i(1-p)^{n-i}.
\]

\vspace{2mm}
\noindent{\bf LRC Codes:}

\noindent The static resilience of a $(n,k)$ LRC code, denoted as $\pi_{L}$, can be
computed as:
\begin{align*}
\pi_{L} =& \Pr\!\left(\!B(n,p)\leq m\!-2\right) + \\
&\Pr\!\left(\!B(n,p)= m\!-1\right)\cdot 2\theta(1-\theta) + \\
&\Pr\!\left(\!B(n,p)= m\right)\cdot(1-\theta)^2.
\end{align*}
The first summand of the expression represents the probability that at most
$n-k-2=m-2$ blocks are inaccessible, and thus, data can always be
reconstructed using the global erasure code. The second summand represents
the probability that $m-1$ blocks are unavailable but one failure can be
repaired using one of the local groups. Given the probability $\theta$ that
at most one block per local coding group is unavailable, \linebreak $\theta =
(k/2+1)p(1-p)^{k/2}$. Then, the coefficient $2\theta(1-\theta)$ represents
the probability of having only one local coding group with at most one
unavailable block. Similarly, the third summand represents the probability
that $m$ blocks are unavailable and at most one block is
unavailable in each local coding group.

\vspace{2mm}
\noindent {\bf CORE's Product Code:}

\noindent Unlike MDS and LRC codes, measuring the static resiliency of
product codes in general is an open problem. Muqaibel~\cite{spcbound} showed
the complexity of measuring it and provided a closed-form expression to
measure the static resiliency of product codes for a specific scenario when
$m=n-k=1$ and $t=1$ for $n\leq 8$, and an upper bound of $\pi_{C}$ for cases
where $n>8$.  However, for the specific class of codes used in CORE (one
dimension uses the simple parity code), a lower bound of the resilience
$\pi_{C}$ can be obtained by considering that reconstruction is feasible only
if there is at most one inaccessible block per column. Thus, $\pi_{C} \geq
\Pr\left( B(n,\vartheta) \leq m \right)$, where $\vartheta = \Pr(B(t+1,p)\leq
1)$ is the probability that there is at most one inaccessible block per
column \cite{spcbound}, which can be rewritten as:
\begin{align*}
\pi_{C} \geq \sum_{i=0}^m \binom{n}{i} \vartheta^i(1-\vartheta)^{n-i},\text{
where}\\
\vartheta = (1-p)^{t+1}+(t+1)p(1-p)^{t}.
\end{align*}

\begin{figure}
  \centering
  \hspace{-6mm}
  \includegraphics[scale=1]{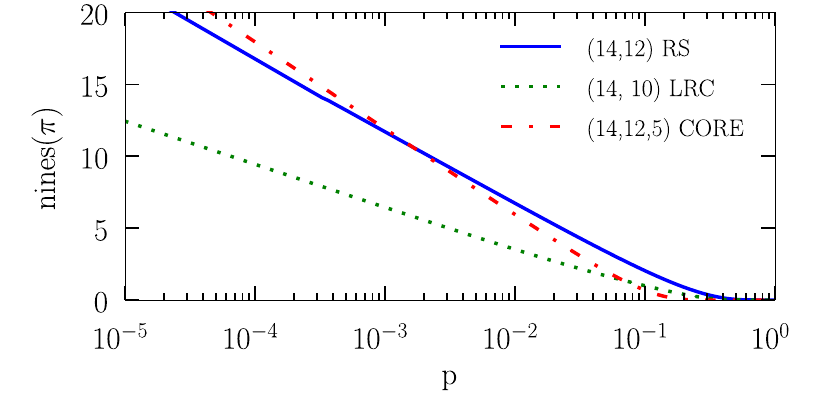}
  \caption{Static resilience $\pi$ (in number of nines) of MDS Reed-Solomon
  code (RS), Local Reconstruction code (LRC) and (lower bound for) CORE's
  product code as a function of the block/node unavailability probability
  $p$. The stretch factor of the RS code is approximately $1.17\times$ whereas for
  LRC and CORE codes is $1.4\times$.}
  \label{f:resil14}
\end{figure}

In Fig.~\ref{f:resil14} we depict the resilience of the three analyzed codes
corresponding to a stretch factor of approximately $1.17\times$ for
the RS code, and of $1.4\times$ for LRC and CORE codes. The static
resilience $\pi$ is often represented using `number of nines', where, e.g.,
$\pi \geq 0.999$ represents a static resilience of three nines (in general,
nines$(\pi)=log_{10}(1-\pi)^{-1}$). We see that for the same storage
overhead as LRC, CORE's product codes achieve much better resilience. Note that the optimality of RS codes (MDS property) allows them to achieve similar resiliency as CORE's product code (ignoring repairability, and the fault tolerance that can be enhanced with repairs) for a lower stretch factor --- $1.17\times$ instead of $1.4\times$.

\subsection{Repair Traffic Requirements}

We next compare the repair traffic requirements of (i) a MDS ($n,k_e$)
erasure code (denoted as \emph{EC}), (ii) a ($n,k_l$) LRC code as described
in Section~\ref{s:background}, and (iii) a ($n,k_c,t$) product code as
described in Section~\ref{s:core1}, with $t$ set to $k_l/2$.  The choice of
cross-object coding parameter $t=k_l/2$ allows a fair comparison between CORE
codes and LRC, since in both cases all the systematic pieces can then be
repaired by contacting exactly $k_l/2$ nodes.

\vspace{2mm}
\noindent {\bf Repairing a Single Block Failure:}

\noindent As discussed in Section~\ref{s:bc}, MDS erasure codes require the
transfer of $k$ blocks to repair even a single failure. In
Section~\ref{s:lrcinazure} we showed that the average repair cost of single
block failures in LRC is reduced to $(2kn-k^2-2k)/2n$. However, in the case
of the CORE's product code, any single failure can be repaired by
transferring only $t$ blocks. For $t=k/2$, the LRC repair cost is larger than
that of CORE, provided than the stretch factor is lower than two.

\vspace{2mm}
\noindent {\bf Repairing Multiple Block Failures:}

\noindent Though repairing a single block failure requires less traffic in
CORE's product codes than in LRC and MDS codes, we cannot use the single
failure model to fairly compare the repair traffic in CORE. Local repairs in
CORE work only when there is at most one failure per encoding group column.
If multiple failures occur in the same column, these failures cannot be
repaired using the ``vertical" parity, increasing the average repair traffic.
For a fairer comparison, we determine the average cost of repairing all
affected data objects when a fraction $p$ of the storage nodes fail.

Consider the random variable $W$ representing the network traffic required to
repair a data object when each of the redundant blocks of the object might
fail with probability $p$. Then, we can compare the repair costs of the
different codes by measuring the conditioned probability $\Pr(W|~\Pi)$, where
$\Pi$ is the event that a given data object can be repaired when a fraction
$p$ of nodes fails.  As discussed in Section~\ref{s:ft}, for large systems,
we can assume that $\Pr(\Pi)=\pi$, where $\pi$ is the code's static
resilience.

In CORE and LRC the traffic required to repair a specific failure pattern
depends on the number of failed blocks, so we express the previous
probability conditioned to the number of failures,
\vspace{-3mm}
\[
\Pr(W\!=\!w|~\Pi) = \sum_{i=0}^n \Pr(W\!=\!w|B(n,p)=i,~\Pi),
\]
To compare the performance of different codes we will measure the expected
value and variance of $W$. Using the laws of the total expectation and the
total variance:
\[
\E(W|~\Pi) = \sum_{i=0}^n \E(W\!=\!w|B(n,p)=i,~\Pi)\cdot\Pr(B(n,p)=i),
\]
\[
\Var(W|~\Pi) = \E(W^2|~\Pi) - \E(W|~\Pi)^2.
\]
For each of the evaluated codes, the expression $\E(W\!=\!w|B(n,p)=i,~\Pi)$ is
measured numerically using a Monte-Carlo experiment, where, at each iteration
$i$ random blocks out of the total $n$ blocks fail. We measure $\E(W^2|~\Pi)$ required
to obtain $\Var(W|~\Pi)$ similarly. We normalize the network traffic values by the size
of the whole stored object --i.e., $kq$ bits.

Similarly, we determine the repair time required to repair each failure
pattern, obtaining $\E(T|~\Pi)$ and $\Var(T|~\Pi)$, where $T$ is distribution
of the repair time. It is measured assuming a congestion free network, but
with end nodes with a limited bandwidth capacity, and considering that the
delays in repair thus occur when a single node sends/receives multiple
blocks. We normalize the repair time $T$ by the time a node requires to
download a whole data object ($k$ blocks) from another node.

\begin{figure}
\hspace{-3mm}
\begin{subfigure}[b]{0.46\textwidth}
  \includegraphics[scale=0.75]{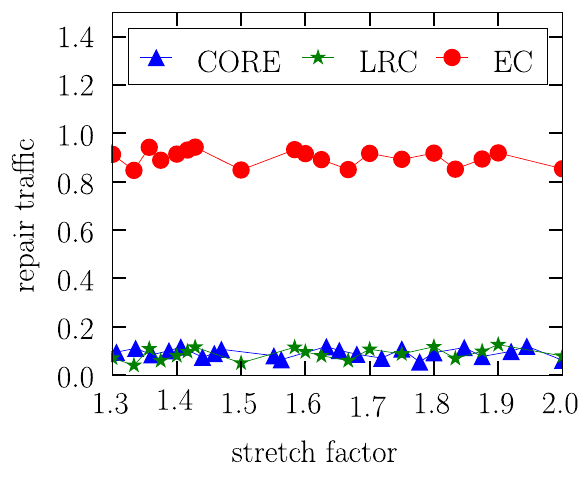}%
  \vspace{-4mm}
  \caption{$p=0.01$}
\end{subfigure}%
\hfill
\begin{subfigure}[b]{0.46\textwidth}
  \includegraphics[scale=0.75]{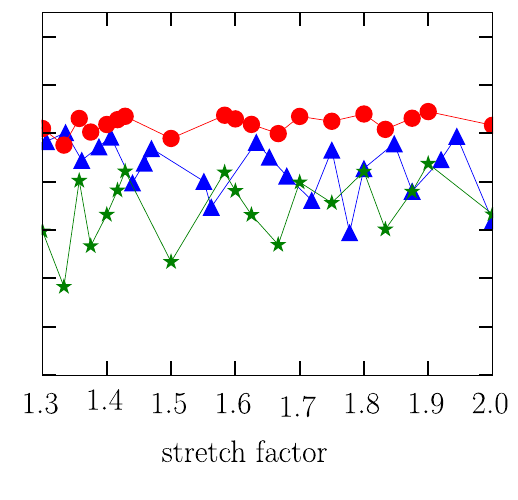}%
  \vspace{-4mm}
  \caption{$p=0.1$}
\end{subfigure}
\caption{Average network traffic (normalized) required to repair a data
object when a fraction $p$ of the storage nodes fails.}
\label{f:failuretraff}
\end{figure}

\vspace{2mm}
\noindent {\bf Multiple Failures Evaluation:}

\noindent {To numerically compare the network traffic $W$ and the repair time
$T$ of each code, we evaluate each metric for different $n$ and $k$ parameter
combinations.  For each stretch factor $n/k$ value, we choose the best
(minimum) network traffic and (minimum) repair time achieved by each code.

Fig.~\ref{f:failuretraff} depicts the average traffic required to repair a
data object when a fraction $p$ of storage nodes fail. This was evaluated for
$p\in\{0.01,0.1\}$, respectively representing concurrent failure of 1\% and
10\% of the nodes. For both low and high failure probabilities, CORE's
product code and LRC require comparable repair traffic (LRC outperforms
slightly). However, from Fig~\ref{f:failuretime} we see how CORE's product
code can reduce the repair time by an order of a magnitude, since
``vertical'' repairs can be concurrently and independently executed. Having
short repair times means that the storage system recover faster from an
unsafe state, which in turn increases the system's robustness. The system is
thus likely to operate in configurations corresponding to low $p$ values.

\begin{figure}
\centering
\hspace{-3mm}
\begin{subfigure}[b]{0.46\textwidth}
  \includegraphics[scale=0.75]{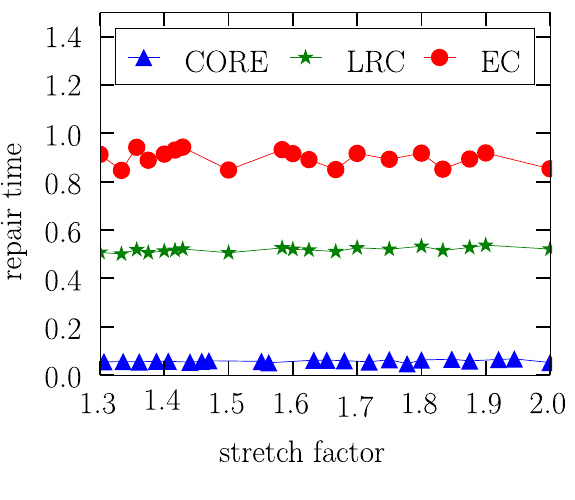}%
  \vspace{-4mm}
  \caption{$p=0.01$}
\end{subfigure}%
\hfill
\begin{subfigure}[b]{0.46\textwidth}
  \includegraphics[scale=0.75]{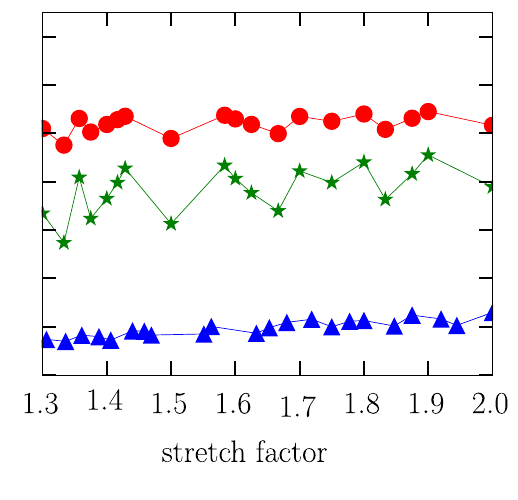}%
  \vspace{-4mm}
  \caption{$p=0.1$}
\end{subfigure}
\caption{Average time (normalized) required to repair a data object when a
fraction $p$ of the storage nodes fails.}
\label{f:failuretime}
\end{figure}

\subsection{Degraded Read Performance}

Finally, we compare the read performance of the different codes when
data is accessed in a degraded system state, i.e., when some storage
nodes and the corresponding data are (temporarily) unavailable,
possibly because of the nodes being overloaded, network outage, or
repair actions being yet to be carried out. We consider two
different typical data access scenarios:

\noindent {\bf Centralized Degraded Read:} A single node in the
system aims to retrieve a whole data object. Such an access pattern
may arise in applications like video-on-demand.

\noindent {\bf Distributed Degraded Read:} The object is read by $k$
nodes, each accessing a different systematic block. Such access
pattern is typical in a MapReduce application where $k$ different
mappers may parse $k$ different blocks.

As earlier, the codes were studied for varying stretch factors
$n/k$. In Fig.~\ref{f:alltraff} we depict the average traffic
required to retrieve a single object in the centralized degraded
read experiments. When the fraction of unavailable nodes is low
($p=0.01$), all three codes can successfully retrieve the stored
object without any additional overhead by transferring only an
amount of data equivalent to the stored object. However, when the
fraction of unavailable nodes is high ($p=0.1$), CORE's product code
has to retrieve some extra data for the low stretch factor
configurations. Low stretch factors in CORE codes impose an extra
read overhead due to the number of extra blocks that have to be
retrieved during the repair of the unavailable systematic blocks. In
Fig.~\ref{f:singletraff} we depict the average read traffic in the
decentralized read experiments, where $k$ distributed processes each
read a different systematic block. All the codes have similar
behavior for low $p$. However for larger $p$, CORE's product codes
achieve similar performance as LRC, and we can see how traditional
erasure codes require slightly more traffic than the other two
codes.

From this macroscopic experiments, we see that for most realistic
system configurations (e.g., code's stretch factor between 1.5-2)
and state (low $p$ values), CORE achieves much better repairability,
while having similar read performance as local reconstruction codes.
Later, in Section \ref{s:exp} we carry out microscopic experiments
with our actual implementation, studying specific fault patterns
within an individual cross-object coded group, to demonstrate
further the advantages of using CORE.

\begin{figure}
\centering
\hspace{-3mm}
\begin{subfigure}[b]{0.46\textwidth}
  \includegraphics[scale=0.75]{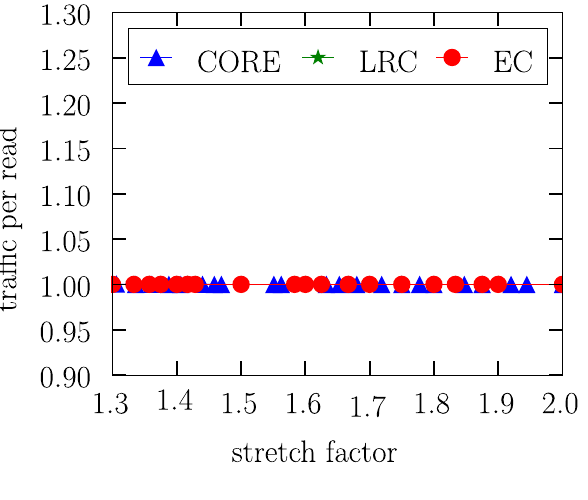}%
  \vspace{-4mm}
  \caption{$p=0.01$}
\end{subfigure}%
\hfill
\begin{subfigure}[b]{0.46\textwidth}
  \includegraphics[scale=0.75]{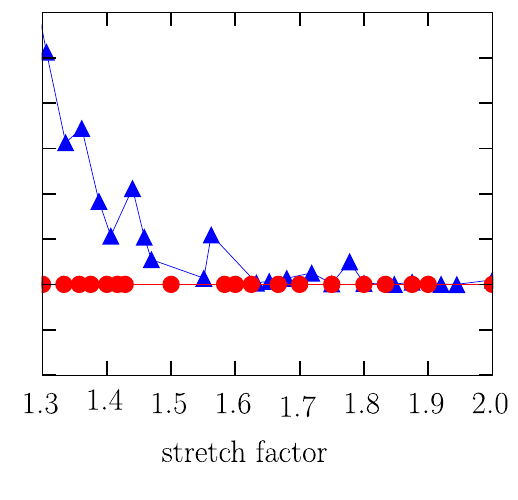}%
  \vspace{-4mm}
  \caption{$p=0.1$}
\end{subfigure}
\caption{Traffic (normalized) required to read a stored object in
the centralized degraded read experiment. We depict the results for
two different fractions $p$ of unavailable nodes.}
\label{f:alltraff}
\end{figure}

\begin{figure}
\centering
\hspace{-3mm}
\begin{subfigure}[b]{0.46\textwidth}
  \includegraphics[scale=0.75]{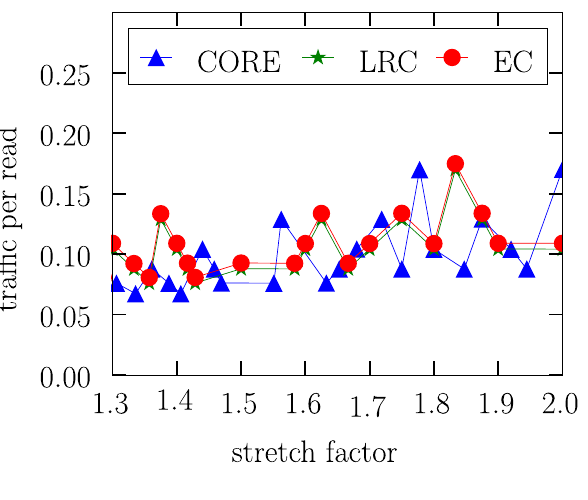}%
  \vspace{-4mm}
  \caption{$p=0.01$}
\end{subfigure}%
\hfill
\begin{subfigure}[b]{0.46\textwidth}
  \includegraphics[scale=0.75]{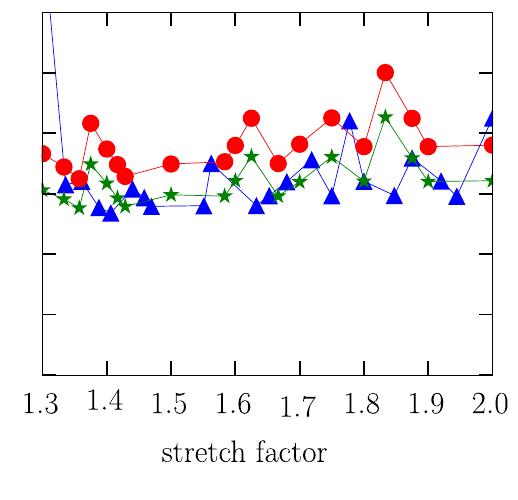}%
  \vspace{-4mm}
  \caption{$p=0.1$}
\end{subfigure}
\caption{Traffic (normalized) required to read a stored object in the
decentralized degraded read experiment. We depict the results for two
different fraction $p$ of unavailable nodes.}
\label{f:singletraff}
\end{figure}

\section{CORE's Algorithmic Aspects}\label{s:algs}
One of the new aspects of CORE --compared to a system of independently-encoded objects-- is its higher level of granularity  which provides new opportunities (e.g., in a CORE scheme of $(n,k,t)$ it is possible to repair an object that has more than $n-k$ failed blocks) and also poses new challenges (e.g., given a pattern of failures, is it possible to \textit{recover} --i.e. repair all the failed blocks -- the CORE matrix? or  what is the best schedule for repairing a set of failures?).

In this section we look at these algorithmic problems and provide solutions for them. 
We adopt a divide-and-conquer approach to tackle these issues.
Specifically, given a matrix representing the available and failed
nodes (subsequently called the CORE matrix), we first split the
failures into `independent clusters' (defined below). Other
algorithms, i.e., recoverability-checking and repair scheduling, can
be performed within each cluster. We discuss these next.

\subsection{Identifying Independent Clusters}
We define disjoint subsets of failed nodes that can be handled
without interference as independent clusters.\footnote{In other
parts of this paper, we also use the term computer/node cluster in
the common sense of the word, which should not be confused with the
failure clusters in the CORE matrix} Essentially, two different
clusters should not share any common row or column containing failed
nodes. Two important benefits of such clusters are (i) they allow
parallel repairs, (ii) they may allow partial recovery when the full
CORE matrix in not recoverable.

A naive way to create the clusters is as follows. Initially, each single failure is considered a cluster. Two clusters are then merged if there exists at least one common row or column on which both clusters have a failure. The process is continued until there are no mergeable clusters left. The number of clusters in a CORE matrix is between 0 and $m$ (number of rows). To investigate the distribution of the number of clusters based on the number of failures, we ran our clustering algorithm on 10M randomly-generated failure matrices  for code parameters (14,12,5) and varied the number of random failures from 1 to 20 (result shown in  Figure~\ref{fig:number_of_clusters}).
\begin{figure}[t]
\begin{center}
    \includegraphics[trim = 1cm 3cm	 2cm 2.5cm, clip, width=\textwidth]{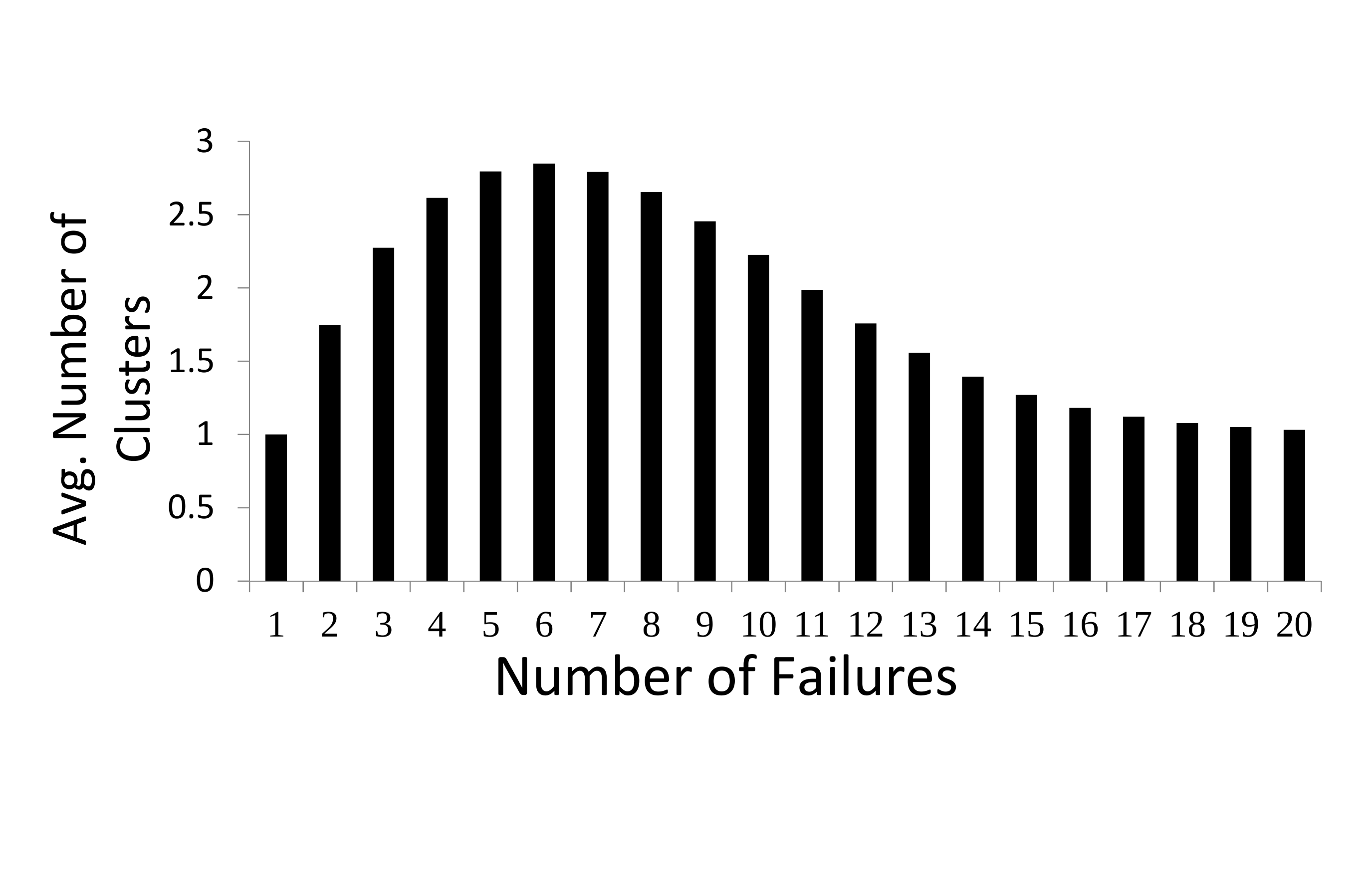}%
		\vspace{-5mm}
    \caption{The average number of clusters versus the number of failures for
    CORE's code parameters (14,12,5)}%
    \label{fig:number_of_clusters}%
\end{center}
\end{figure}

\subsection{Recoverability-Checking Algorithm}

For coding schemes that work at the level of single objects, given the set of failures, one can directly infer whether an object is recoverable. In the case of CORE, however, this is more subtle. For instance, objects may still be recoverable even if there are more than $n-k$ failed blocks within a single CORE row. We first identify two bounds and
then introduce an algorithm to determine an object's recoverability.\newline

\noindent\textbf{The (Ir)Recoverability Bounds}. For a $(n,k,t)$ code:\\
\begin{itemize}
\item the \textit{lower} bound of \textit{irrecoverability}, $L$, is:
$$2 \times (n-k+1)$$
It occurs if two\footnote{Any single-row failure pattern is
always recoverable.} rows are minimally irrecoverable (each has $n-k+1$
failures) and the column indexes of their failures are identical
(i.e., no vertical repair possible).

\item the \textit{upper} bound of \textit{recoverability}, $U$, is:
$$t \times (n-k) + (2k-n) \times 1$$
This occurs when all rows are maximally recoverable (each has $n-k$
failures) and have identical failure column indexes (i.e., the
remaining $k-(n-k)=2k-n$ columns can each tolerate a single
failure).
\end{itemize}

These two bounds define an interval. For any failure number outside of this interval, the ir/recoverability can be immediately decided. More precisely, if the number of failures is smaller than $L$ then the pattern is recoverable -- although, as we will see later, this is a very pessimistic bound -- likewise, if the number is greater than $U$, then it is certainly not recoverable. 

For all the values within the above interval (inclusive), the outcome depends on the distribution of the failures. We propose a recursive algorithm which is able to decide whether a given CORE matrix with a specific failure pattern is recoverable or not. At each step of the algorithm, all the repaired and repairable rows/columns are removed and the algorithm restarts with the reduced matrix as the new input. If it results in an empty matrix, then the patterns is recoverable, otherwise it is not.

We implemented this algorithm and used it to carry out an analysis
on the recoverability likelihood of different patterns.
Figure~\ref{fig:rec_like}, obtained from 10M random runs, shows the
recoverability likelihood (in terms of number of 9's) of the CORE matrix of size (14,12,5) for
all possibly recoverable failure numbers ($\le$ $U$ = 20). It
clearly illustrates the fact that CORE's lower bound of
irrecoverability ($L$ = 6, in this setting) is too strict.
\begin{figure}[t]
\begin{center}
    \includegraphics[trim = 2cm 2cm 2cm 2cm, clip, width=\textwidth]{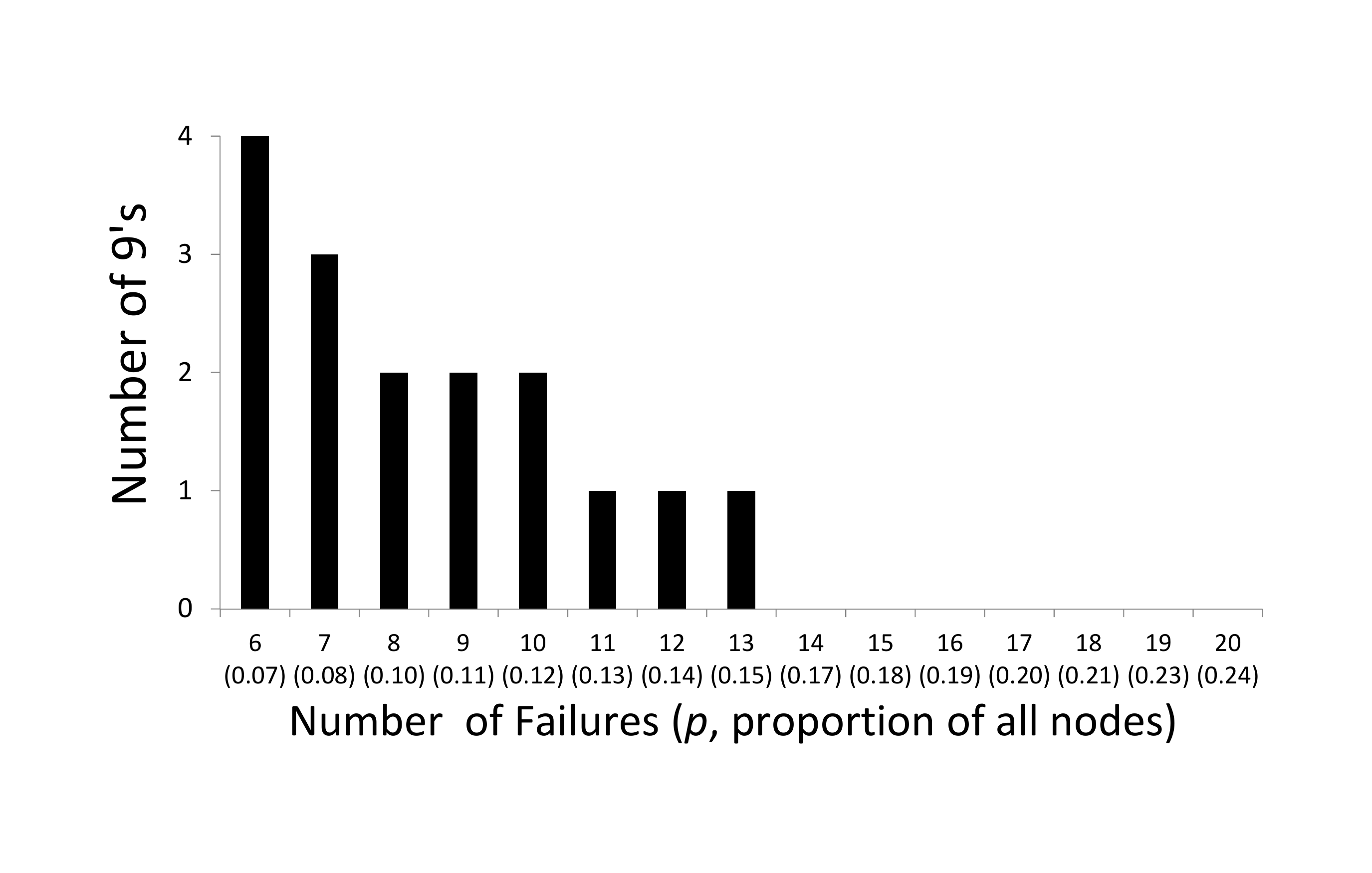}%
		\vspace{-5mm}
    \caption{The recoverability likelihood  of the scheme (14,12,5) in terms
    of number of 9's based on the number of failures (or equivalently,
    $p$).}%
    \label{fig:rec_like}%
\end{center}
\end{figure}

\subsection{Repair Scheduling Algorithms}\label{s:scheduling-algs}
Many different repair schedules may exist for a given fault pattern. Here, we first investigate two straw man approaches, namely \textit{column-first} and \textit{row-first}, then propose an algorithm called Recursively Generated Schedule (RGS). Analytical and experimental studies show that RGS outperforms the baseline approaches.

The \textit{column-first} algorithm always gives higher priority to vertical repairs and applies horizontal repair when no further vertical repairs are possible. The \textit{row-first} analogously prefers horizontal repairs. In both algorithms, while doing horizontal repairs, always the best candidate (the one with maximum number of failures but still repairable) is prioritized over the other ones.\newline

\noindent\textbf{Recursively Generated Schedule (RGS) algorithm:}. This algorithm first identifies the critical set of failures (failures that decrease the minimum number of required vertical or horizontal repairs) and repairs them first, along the call chain of a recursive cost function $c$. All other repairs (non-critical ones) are scheduled using $c'$, a non-recursive cost function.\newline

\noindent In order to identify the critical failures, we define two variables, $v$ and $h$, as follows:
$$v = \sum_{i = 1}^{t} minV(Row_i) ~~~~~~;~~~~~ h = \sum_{j = 1}^{k} minH(Col_j)$$

\noindent in which, $minV(Row_i)$ returns the minimum number of \textit{vertical} repairs required by row $Row_i$, and $minH(Col_j)$ returns the minimum number of \textit{horizontal} repairs  required by column $Col_j$, more precisely:
$$minV(Row_i) = \left\{ \begin{array}{l l}  0 & \quad \textrm{if $|X| \le (n -k) $ }\\    |X| - (n -k) & \quad \textrm {otherwise}   \end{array} \right.$$
$$minH(Col_j) = \left\{ \begin{array}{l l}  0 & \quad \textrm{if $|X| \le 1 $ }\\    |X| - 1 & \quad \textrm {otherwise}   \end{array} \right.$$

\noindent The most important element of RGS is the recursive cost function $c(h,v)$ defined as:
$$ c(h,v) = \left\{ \begin{array}{l l}    c(h,dec(v)) + t & \quad \textrm{if $v > 0$}\\    & \\ c(dec(h),v) + k & \quad \textrm{if $v = 0$} \\ & \textrm{or $dec(v)$ is not applicable}   \end{array} \right.$$
in which $dec(v)$ and  $dec(h)$ reflect the decreases in the values of $v$ and $h$ after a single repair is performed.

The cost function $c$ decreases the values of first $v$  and then $h$ by at least one unit at each recursion step until we reach $c(0,0)$, which is the \textit{base case}\footnote{If the failure pattern is recoverable, then $c(h,v)$ will always reach the base case.}.
The notable property of the base case is that any remaining repair can be done either vertically or horizontally. In other words, there is at most one failure per column, and at most $n-k$ failures per row. Therefore, all remaining repair decisions can be safely made using the static cost function $c'$ defined below:
$$ c'(r) = \left\{ \begin{array}{l l}    k & \quad \textrm{if repaired horizontally }\\    r\times t& \quad \textrm {if repaired vertically}   \end{array} \right.$$
in which $r$ denotes the number of remaining repairs for a given row.

\begin{table*}[ht]
\begin{center}
\resizebox{0.95\columnwidth}{!}{%
\begin{tabular}{c|c|c|c|c}
\multicolumn{1}{l}{} & & \textbf{Row-First} & \textbf{Column-First} & \textbf{RGS} \\  \hline \hline 
\multirow{2}{*}{
\rotatebox{90}{\textbf{Step}}}&\textit{Schedule} & $R_3, R_2$ & $C_1,R_2, C_0$ & $c(1,0) \overset{R_3 }\rightarrow c(0,0) \rightarrow C_1$   \\ \cline{2-5}
                          & \textit{Cost} & $2k=24$ & $2t+k=22$ &$k+t=17$ \\ \hline \hline 
\multirow{2}{*}{
\rotatebox{90}{\textbf{Plus}}}&\textit{Schedule} & $R_1, R_3, C_0, R_2$ & $C_0, C_2, R_1, R_2, C_1$ & $c(2,1) \overset{C_0}{\rightarrow}  c(2,0) \overset{R_2}{\rightarrow}$\newline $\overset{R_2}{\rightarrow}c(1,0)\! \overset{R_1}{\rightarrow}\!  c(0,0)\! \rightarrow \!C_1$ \\ \cline{2-5}
                          & \textit{Cost} & $3k+t=41$ & $3t+2k=39$ & $2t+2k=34$ \\ \hline
\end{tabular}}
\vspace{-3mm}
\caption{The analytical cost (number of blocks read ) of repairing the Step and Plus failure patterns using Row-First, Column-First, and RGS algorithms where $k=12$ and $t=5$.}
\label{tab:example-v-vi}
\end{center}
\end{table*}

To demonstrate the differences between the repair schedules
generated by the above three algorithms, we use two failure pattern
examples in the CORE matrix of size (14,12,5): a 3-failure
\textit{step}-shaped pattern  and a 5-failure \textit{plus}-shaped
one. These examples are shown below:\newline
$$\begin{pmatrix}
             \dots & 0 & 0 & \dots \\[0.3em]
             \dots & 0  & 0 & \dots \\[0.3em]
             \dots & \textrm{\textbf{X}} & 0 & \dots \\[0.3em]
       \dots & \textrm{\textbf{X}}  & \textrm{\textbf{X}} & \dots \\[0.3em]
       \dots & 0 & 0 & \dots
     \end{pmatrix}    
,
\begin{pmatrix}
                \dots & 0   & 0 & 0 & \dots \\[0.3em]
       \dots & 0 & \textrm{\textbf{X}} & 0 & \dots \\[0.3em]
       \dots & \textrm{\textbf{X}} & \textrm{\textbf{X}}    & \textrm{\textbf{X}} & \dots \\[0.3em]
       \dots & 0 & \textrm{\textbf{X}}  & 0 & \dots \\[0.3em]
       \dots & 0    & 0 & 0 & \dots
     \end{pmatrix}
$$\newline

It should be noted that since swapping any two rows or any two columns
in the CORE matrix results in an equivalent failure matrix, each of
these patterns represents a class of failure patterns and not
singular instances. Table~\ref{tab:example-v-vi} presents the schedules generated  by 
each algorithm for each failure pattern along with its calculated cost in terms of repair traffic. The corresponding experimental results
are reported in Section~\ref{s:exp}.

Finally, we generalized our analytical study of the above three algorithms to include failure patterns of size 1 to 20. The results  for 10,000 randomly-generated recoverable failure patterns are depicted in Figure~\ref{fig:rec-algs-simluation}. 
Four conclusions can be drawn from this figure: (i) RGS and column-first perform better than row-first and this is especially noticeable when the number of failures is very small (which is, in essence, the MDS code vs. CORE comparison); (ii) as the number of failures and consequently the number of choices to make increases, the benefits of RGS over column-first become more pronounced; (iii) for the large failure numbers, distinct schedule possibilities are limited, and all the algorithms perform similarly; and finally (iv) a more general conclusion is that if one wishes to avoid the relatively complex scheduling algorithms, then the naive column-first approach nevertheless delivers significant benefits w.r.to the row-first (which is roughly like for MDS codes), highlighting the immediate benefits of CORE's product code.\newline\newline \newline

\begin{figure}[t]
\begin{center}
    \includegraphics[trim = 1cm 2cm 5cm 3cm, clip, width=\textwidth]{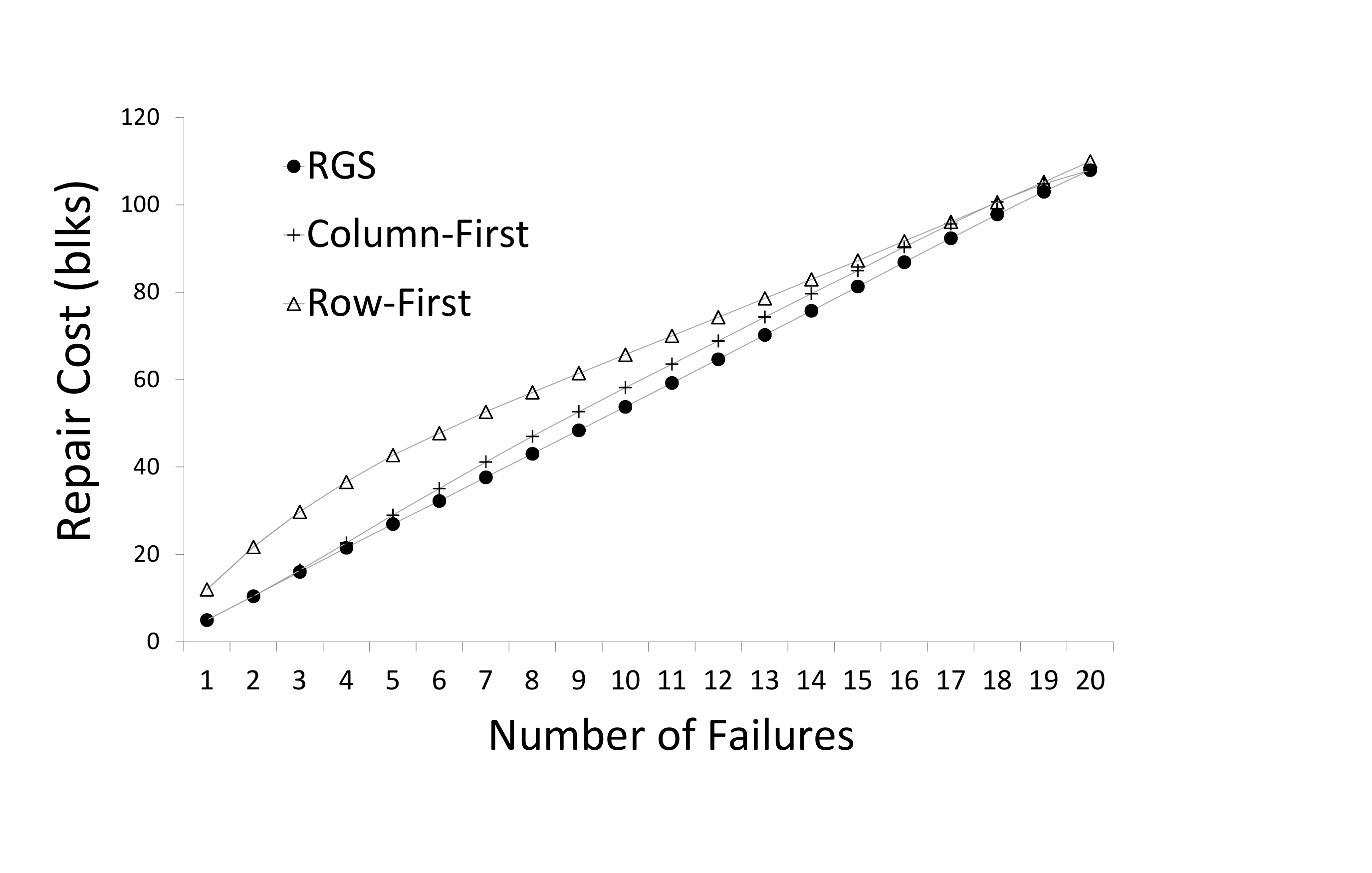}
		\vspace{-5mm}
    \caption{Comparing the Column-First, Row-First, and RGS  algorithms w.r.t
    number of blocks required to carry out the repair.}
    \label{fig:rec-algs-simluation}
\end{center}
\end{figure}

\begin{figure}[ht]
    \begin{subfigure}[b]{\textwidth}
        \includegraphics[trim = 1cm 2.2cm 2cm 2.2cm, clip, width=\textwidth]{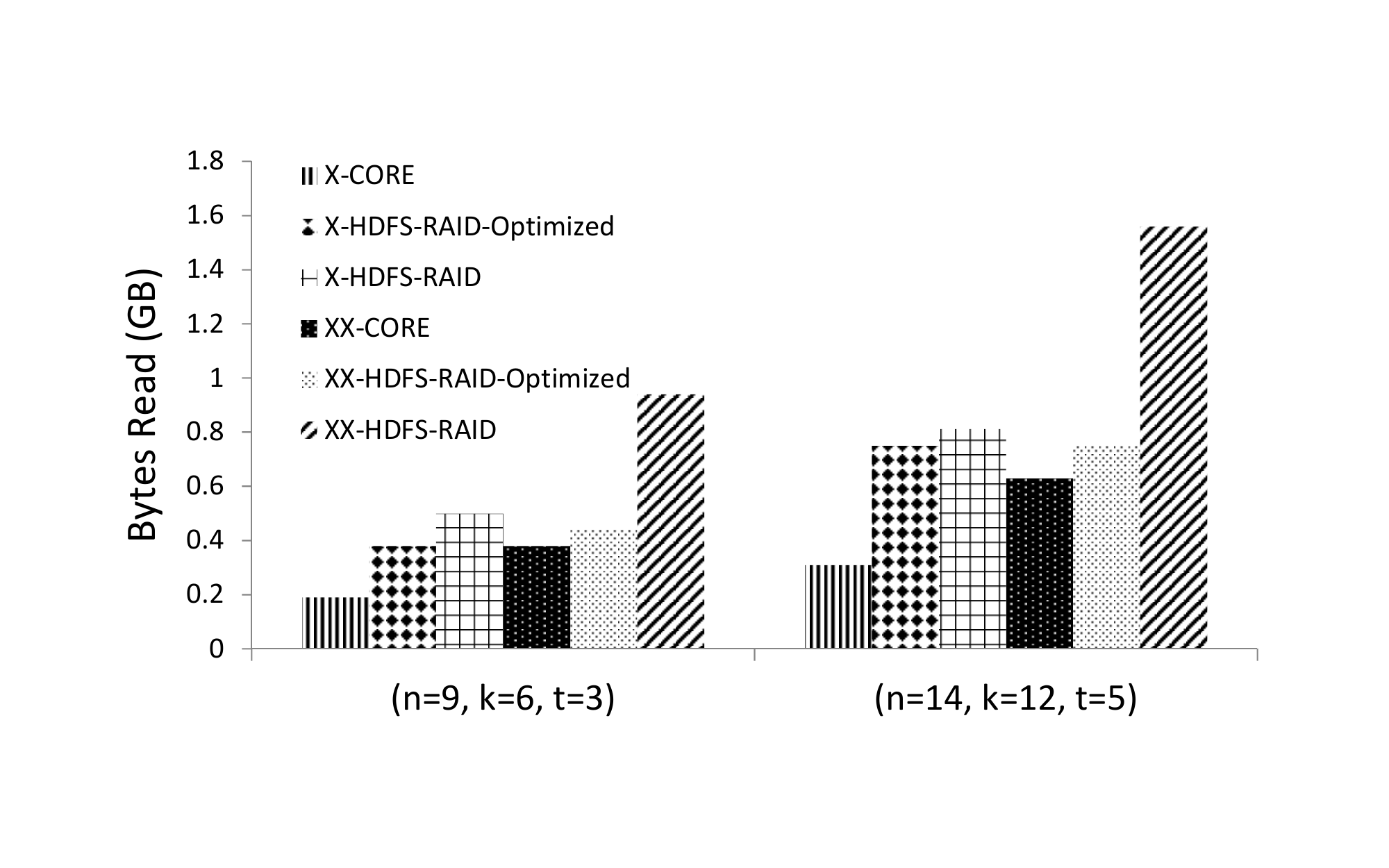}
        \caption{Transferred data}
    \end{subfigure}
    
    \begin{subfigure}[b]{\textwidth}
        \includegraphics[trim = 0cm 2.2cm 3cm 1cm, clip, width=\textwidth]{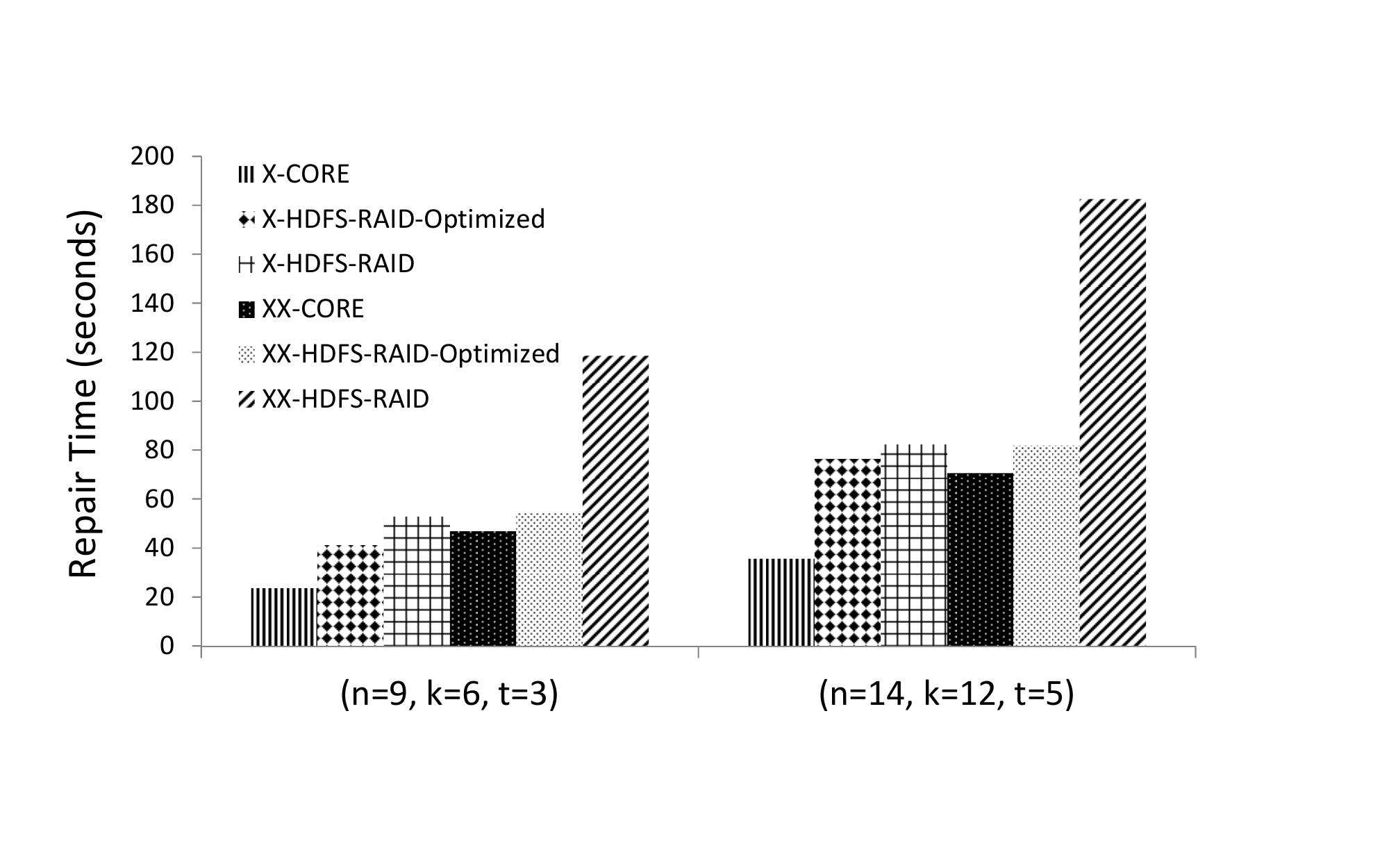}
        \caption{Time (network-critical cluster)}
    \end{subfigure}
    
    \begin{subfigure}[b]{\textwidth}
        \includegraphics[trim = 1cm 2.2cm 2cm 1cm, clip, width=\textwidth]{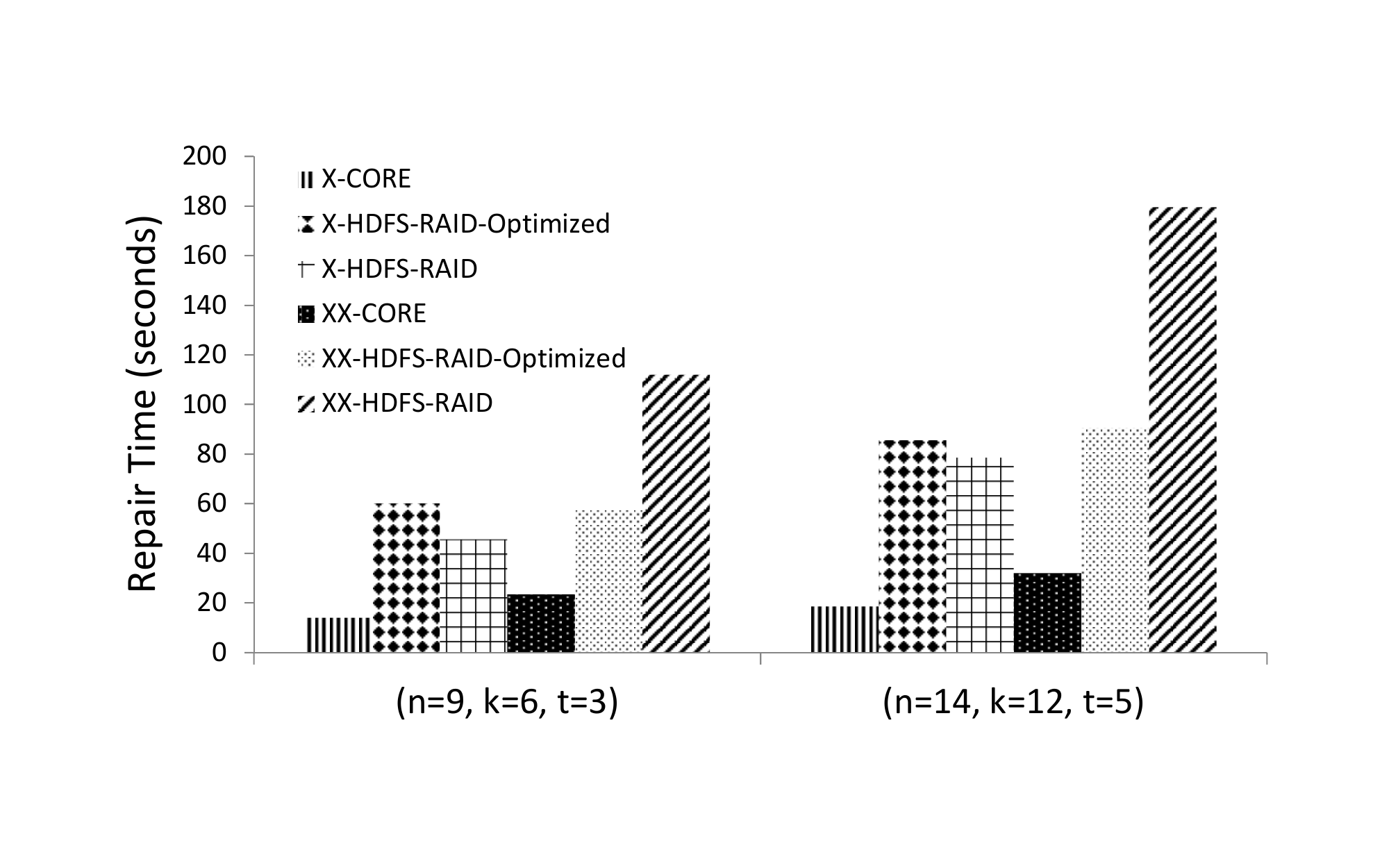}
        \caption{Time (computation-critical cluster)}
    \end{subfigure}		
    \caption{Comparing the repair performance of HDFS-RAID, HDFS-RAID-Optimized, and CORE}
    \label{fig:repair-time-bytes}
\end{figure}

\section{Implementation}\label{s:impl}
To implement the CORE primitive, we used HDFS-RAID~\cite{hdfsraid}, an open-source module inspired by DiskReduce~\cite{diskreduce2}, and developed at Facebook. It wraps around Apache Hadoop's distributed file system (HDFS) and provides HDFS with basic erasure coding capabilities (encoding and decoding). Below, we first introduce HDFS-RAID, then explain two optimizations that we did on HDFS-RAID to improve its performance, and finally give an overview of our implementation of CORE.

\subsection{HDFS-RAID}
HDFS-RAID embeds the Apache HDFS inside an erasure code-supporting wrapper file system named Distributed Raid File System (DRFS). DRFS supports both Reed-Solomon coding as well as simple XOR parity files. These two coding alternatives are orthogonal and used separately based on user preference. Furthermore, both provide two basic features: encoding (a.k.a RAIDing) data blocks and repairing the corrupt/missing blocks.

The two main components  of HDFS-RAID are RaidNode and BlockFixer. RaidNode is a daemon responsible for the creation and maintenance of parity files for all data files. 
Since the default block policy of HDFS is not aware of the dependency relation between the data and parity blocks of a given file, HDFS-RAID manages the placement of parity blocks to avoid co-location of data blocks  and parity blocks. The BlockFixer component reconstructs missing or corrupt blocks by retrieving the necessary blocks, encoding/decoding them, and sending the reconstructed blocks to new hosts.\newline

\subsection{HDFS-RAID Optimizations}\label{s:opts}
In our experiments with HDFS-RAID, we noticed two common performance inefficiencies, and optimized them: \newline

\textbf{Opt1:} The HDFS-RAID implementation uses the \textit{generator polynomial} (and not the more well-known \textit{generator matrix}~\cite{hall2003notes}) representation of Reed-Solomon codes.  In this representation, typically and as is in the HDFS-RAID implementation, always \textit{all} the remaining blocks of a given stripe (which can be more than $k$) are used to repair the missing ones. Generally, this use of extra blocks results in faster decoding, since there will be fewer equations to solve. However, for cases in which network is a bottleneck, this trade-off (fetching extra blocks versus faster decoding) does not pay off. Our optimized version retrieves exactly \textit{k} blocks and ``pretends'' that all other $n-k$ blocks are missing. As confirmed by our experimental results, the bandwidth-scarce clusters can greatly benefit from this optimization. 
\newline

\textbf{Opt2:} The HDFS-RAID implementation implicitly assumes that there is only a single failure per stripe. In case there are more failures, they are discovered only when the read access attempts fail. These newly-detected failed blocks are then added to the list of failed blocks, and the repair process starts again. Our optimized implementation checks for multiple failures beforehand, and repairs them simultaneously, amortizing the repair costs.


\subsection{CORE Implementation}
The CORE storage primitive has been organically integrated with HDFS-RAID by extending the two main functionalities as described below. Since all changes have been made within the RAID subdirectory of the HDFS's code, replacing the corresponding Java library is sufficient to upgrade HDFS-RAID to CORE.

\textbf{RAIDing:} The CORE implementation allows vertical coding across files in a given directory. The cross-object stripe size parameter can be configured similar to the stripe-size of HDFS-RAID. Then vertical encoding is reused in the full matrix RAIDing (first row-by-row, then column-by-column, for both data and parity blocks).

\textbf{Repair:} An additional vertical repair option is introduced. The 2-dimensional repair feature implements all the algorithms discussed in Section~\ref{s:algs}: (i) failure detection and failure matrix population, (ii) failure clustering, (iii) recoverability-checking, and (iv) repair scheduling.
	
The correctness of our implementation was verified through multiple test cases in which the MD5 hash values of the repaired files were compared against those of the original files. The source codes, binary distribution, and documentations of our implementation are available at \textcolor{blue}{\url{http://sands.sce.ntu.edu.sg/StorageCORE}}.

\section{Experiments}\label{s:exp}
We benchmarked the implementation with experiments
run on two different HDFS clusters of 20 nodes each:\\
$\bullet$ \textbf{Network-Critical} cluster: A university cluster
which has one powerful PC (4$\times$3.2GHz  Xeon Processors with 4GB of RAM) hosting the NameNode/RaidNode and 19 HP t5745 ThinClients acting as DataNodes.  The average bandwidth  of this cluster is 12MB/s.
\\
$\bullet$ \textbf{Computation-Critical} cluster: An Amazon EC2 cluster
of 20 homogeneous nodes of type \textit{m1.small} (approximately,  1.2 GHz 2007 Xeon Processor with 1.8GB of RAM). In this cluster one node is
hosting the NameNode/RaidNode and the rest are used as DataNodes. The maximum bandwidth between EC2 \textit{m1.small} instances is 250MB/s.

The block size ($q$) used was 64MB. Files were added to HDFS and encoded horizontally first, and then the vertical parity was computed.

We ran two sets of experiments, a first set to compare the
performance of CORE with that of HDFS-RAID, and a second set to
study the repair scheduling algorithms. In both sets, we primarily use the
\textbf{completion time} of the repair process as the main
comparison measure. However, we also measured the amount of
\textbf{transferred data}  in each experiment (as repair traffic). The data transfer numbers serve
two purposes: (i) to verify the correctness of our implementation
--they must match the analytical numbers--  and (ii) to use as a reference point in
analyzing the completion time numbers -- since the amount of transferred data is independent of the type of cluster used. }

Finally, in all experiments the
reported numbers are the average of 10 runs. Since the
variations were small (up to few percents), they are omitted from
the graphs.

\begin{figure*}
\begin{subfigure}[ht]{0.32\textwidth}
  \includegraphics[trim = 1cm 2cm 4cm 2cm, clip, width=\columnwidth]{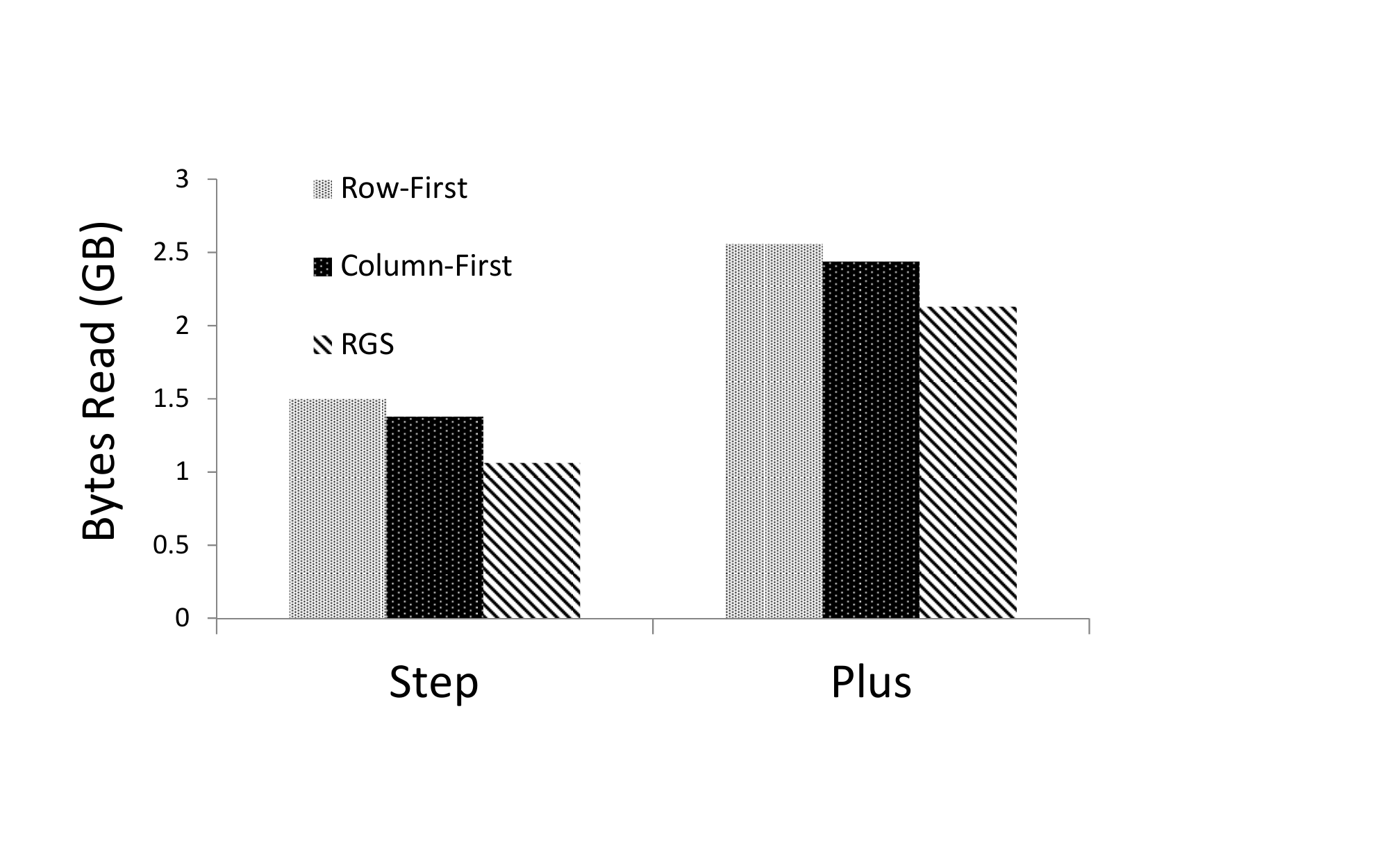}
  \caption{Transferred data}
\end{subfigure}
\begin{subfigure}[ht]{0.33\textwidth}
  \includegraphics[trim = 1cm 2cm 4cm 2cm, clip, width=\columnwidth]{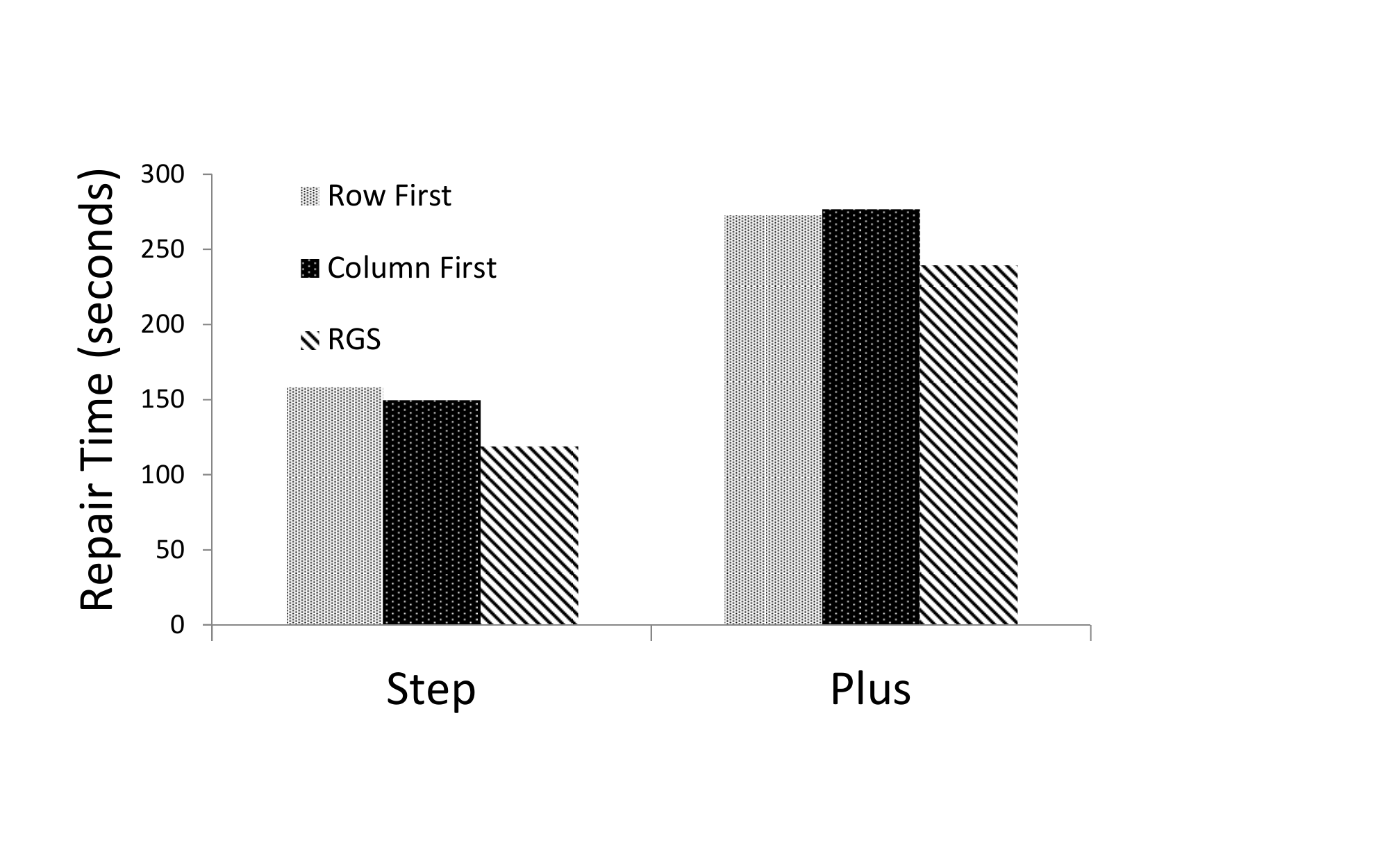}
  \caption{Time (network-critical cluster)}
\end{subfigure}
\begin{subfigure}[ht]{0.33\textwidth}
  \includegraphics[trim = 1cm 2cm 5cm 2cm, clip, width=\columnwidth]{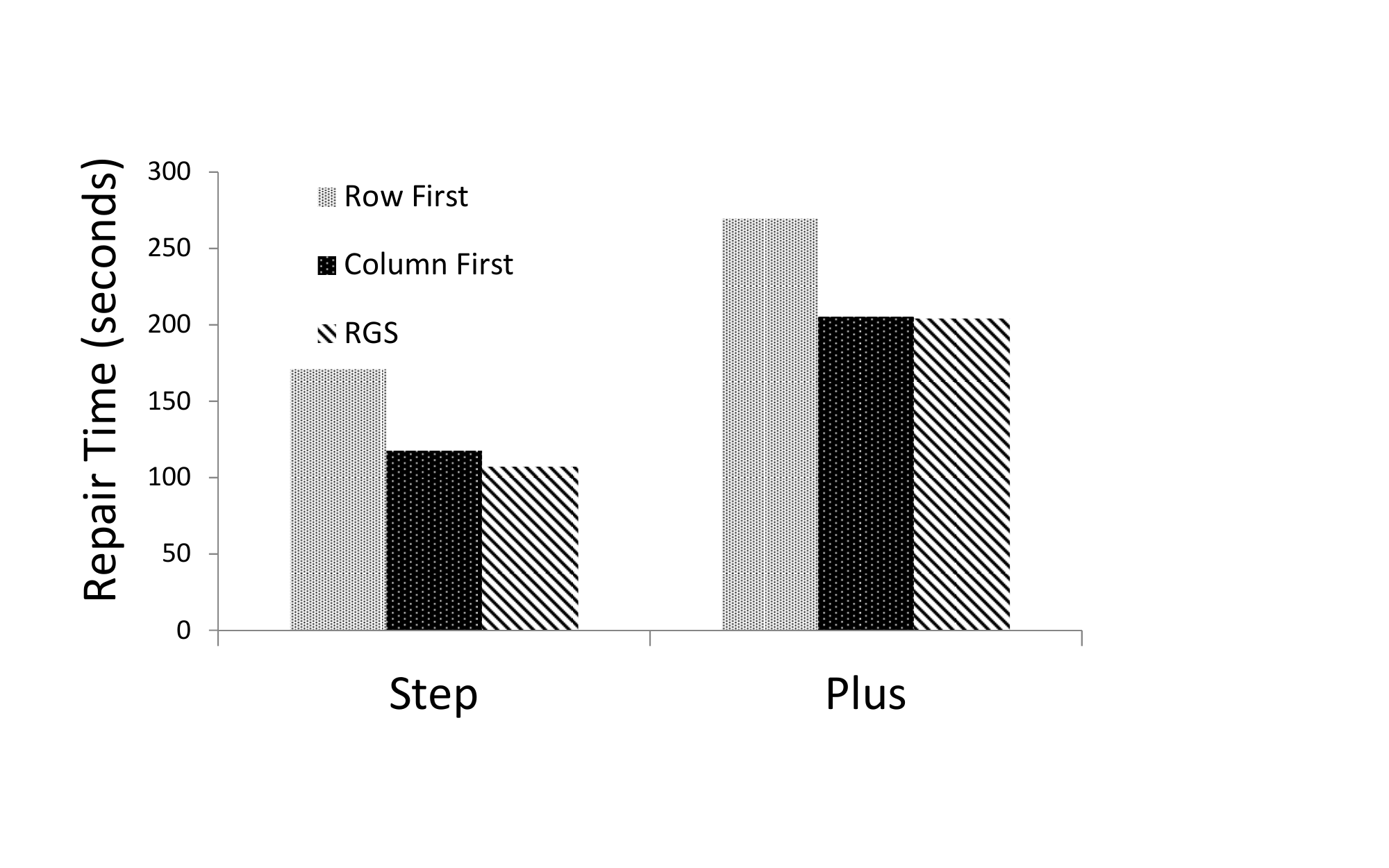}
  \caption{Time (computation-critical cluster)}
\end{subfigure}
\caption{Performances of the repair scheduling algorithms on two
different failure patterns.} \label{fig:rec-algs}
\end{figure*}

\subsection{CORE vs. HDFS-RAID}
In these experiments, we compared three methods (namely, HDFS-RAID,
HDFS-RAID-Optimized and CORE) using two different sets of coding
parameters: (9,6,3) and (14,12,5), inspired respectively by the code length and storage overheads of Google's GFS and Microsoft Azure. In each case two different
failure patterns were enforced: a one-failure pattern represented by
\textbf{X} and a two-failures pattern represented by \textbf{XX}. For the
two-failures pattern, both are set to happen in the same object
(i.e., on the row). The reason for this setting is two-fold: (i) it
favors the HDFS-RAID since at almost the same cost it can repair two
failures instead of one; (ii) if two failures happen on different
rows, the experiment will be, in effect, a variation of the
one-failure pattern.

From the results shown in Figure~\ref{fig:repair-time-bytes}, we can
draw several conclusions:
\begin{itemize}
\item For single failure, the overhead of CORE is less than 50\%
of HDFS-RAID. This is due to the two inherent advantages of CORE: (i) single failure can be repaired
vertically, using far fewer blocks, and (ii) it uses a much cheaper
XOR operation instead of expensive decoding/re-encoding (this is
particularly significant in the computation-critical
cluster).
\item The impact of our first HDFS-RAID optimization (Opt1 in
Section~\ref{s:opts}) can be seen in the results (the difference
between the 2nd and the 3rd chart bars). As explained before, this
optimization  is targeted \textit{specifically} for the clusters in which
network is a scarce resource (part $b$ in Figure~\ref{fig:repair-time-bytes}). The improvements are particularly
pronounced in cases where the number of avoided block retrievals are
higher (e.g., one failure in the scheme (9,6,3)).
\item The gains from our second HDFS-RAID optimization (Op2 in
Section~\ref{s:opts}) are also noticeable (the 5th and the 6th chart
bars in all setups). 
\item Growth in the CORE matrix size, from (9,6,3) to (14,12,5), results in even higher
gains, especially in clusters where computation power is scarce.	
\end{itemize}

\subsection{Repair Scheduling Algorithms}
In this set of experiments, the three repair scheduling algorithms
of Section~\ref{s:scheduling-algs} were compared using the
\textit{Step} and \textit{Plus} failure patterns. HDFS-RAID has
neither a notion of repair scheduling -- it treats objects
independently -- nor can it fully recover from the Plus failure
pattern, so it was not considered in the following experiments.

These experiments were run for CORE matrix of size (14,12,5). The
results are shown in Figure~\ref{fig:rec-algs} and  as expected, the data part of this figure (part \textit{a}) mirrors the analytical results presented in Table~\ref{tab:example-v-vi}. Moreover, the completion time numbers (parts \textit{b} and \textit{c}) are also, to large extent, in-line with the data results. The only two discrepancies are explained below:
\begin{itemize}
\item Completion time of the Column-First algorithm on the Plus pattern in the network-critical cluster (part \textit{b}) is longer than expected. This is caused by the last repair 
which uses two other freshly-repaired blocks. Accessing those blocks is delayed until NameNode's heartbeat-driven mapping tables are updated.

\item Completion time of the RGS algorithm in the computation-critical cluster (part
\textit{c}) is only slightly better than that of Column-First, despite applying one vertical repair less (see Table~\ref{tab:example-v-vi} for the schedules). This is due to the fact that for these patterns the RGS and Column-First apply the same number of horizontal repairs and these are the main driving factor of the cost in the computation-critical
cluster.

\end{itemize}

\section{Conclusions \& Future Work}

In this paper we demonstrated that some simple and standard techniques (and thus easy to implement and organically integrate) can provide significant data repair and access boost in erasure coded distributed storage systems. We studied our approach of introducing cross-object coding on top of normal erasure coding analytically, comparing it with both traditional MDS codes as well as very recently proposed Local Reconstruction Codes (used in Azure). The ideas were implemented (as the CORE storage primitive) and integrated organically with HDFS-RAID, and benchmarked over a proprietary cluster and EC2. Analytical \& numerical studies, as well as experiments with the real implementation all demonstrate the superior performance of CORE over state-of-the-art techniques for data reads and repairs. While naive solutions can be readily used, in future we will like to explore the CORE code properties to achieve better performance also during data insertion/updates. The current evaluations are static, based on snapshots of the system state. We speculate that CORE's better repair properties will yield a system in a better state over time. We will thus carry out trace driven experiments to study the system's dynamics better.

\balance
\bibliographystyle{acm}
\bibliography{CORE}

\end{document}